\documentclass[two_column]{aa}  
\usepackage{graphicx}
\usepackage{float}
\usepackage{multirow}
\usepackage{siunitx}
\usepackage{color}
\usepackage{verbatim}
\usepackage{txfonts}
\begin{document} 

   \title{Impact of binary interaction on the evolution of blue supergiants}
   \subtitle{The flux-weighted gravity luminosity relationship and extragalactic distance determinations}
   %\title{Flux-weighted gravity-luminosity relationship: Impact of close binary evolution}

   \author{E. J. Farrell\inst{\ref{tcd},\ref{unige}} 
   \and J. H. Groh\inst{\ref{tcd}}
   \and G. Meynet\inst{\ref{unige}}
   \and R. Kudritzki\inst{\ref{hawaii},\ref{munich}}
   \and J.J. Eldridge\inst{\ref{auckland}}
   \and C. Georgy\inst{\ref{unige}}
   \and S. Ekstr{\"o}m\inst{\ref{unige}}
   \and S.-C. Yoon\inst{\ref{korea}}
   }  

   \institute{School of Physics, Trinity College Dublin, The University of Dublin, Dublin, Ireland\label{tcd}%\\
              %\email{tcd email}
         \and
             Geneva Observatory, University of Geneva, Chemin des Maillettes 51, 1290 Sauverny, Switzerland\label{unige}%\\
             %\email{email}
         \and
             Institute for Astronomy, Univeristy of Hawaii, 2680 Woodlawn Drive, Honolulu, HI 96822\label{hawaii}%\\
             %\email{email}
         \and
             University Observatory Munich, Scheinerstr. 1, D-81679 Munich, Germany\label{munich}%\\
             %\email{email}
         \and
             Department of Physics, Private Bag 92019, University of Auckland, New Zealand\label{auckland}%\\
             %\email{email}
         \and
                Department of Physics and Astronomy, Seoul National University, Gwanak-ro 1, Gwanak-gu, Seoul, 08826, Korea\label{korea}%\\
             }

   \date{Received ; accepted }

% \abstract{}{}{}{}{} 
% 5 {} token are mandatory

 \abstract{A large fraction of massive stars evolve in interacting binary systems, which dramatically modifies the outcome of stellar evolution. We investigated the properties of blue supergiants in binary systems and whether they are suitable for extragalactic distance determinations using the flux-weighted gravity luminosity relationship (FGLR). This is a relationship between the absolute bolometric magnitude $M_{\rm bol}$ and the spectroscopically determined flux-weighted gravity $g_{\rm F} = g/T^4_{\rm eff}$, where $g$ is the surface gravity and $T_{\rm eff}$ is the effective temperature. We computed a grid of binary stellar evolution models with MESA and use the v2.1 BPASS models to examine whether they are compatible with the relatively small scatter shown by the observed relationship. Our models have initial primary masses of 9 -- 30 $M_\odot$, initial orbital periods of 10 -- 2511 days, mass ratio $q$ = 0.9, and metallicity $Z$ = 0.02. We find that the majority of primary stars that produce blue supergiant stages are consistent with the observed FGLR, with a small offset towards brighter bolometric magnitudes. In between 1\% -- 24\% of cases, binary evolution may produce blue supergiants after a mass transfer episode, that lie below the observed FGLR. A very small number of such stars have been found in extragalactic FGLR studies, suggesting that they may have evolved through binary interaction. Some models with shorter periods could resemble blue hypergiants and luminous blue variables. We used CMFGEN radiative transfer models to investigate the effects of unresolved secondaries on diagnostics for $T_{\rm eff}$ and $g$, and the biases on the determination of interstellar reddening and $M_{\rm bol}$. We find that the effects are small and within the observed scatter, but could lead to a small overestimate of the luminosity, of $T_{\rm eff}$ and of $g$ for extreme cases. We conclude that the observed flux-weighted gravity luminosity relationship can, in principle, be well reproduced by close binary evolution models. We outline directions for future work, including rotation and binary population synthesis techniques.}


   \maketitle
%
%________________________________________________________________

\section{Introduction} \label{sec:intro}
Massive stars have significant effects on the evolution of their host galaxies, as they are among the most important sources of ionising photons and producers of many of the chemical elements. 
The vast majority of O and B-type massive stars with masses 8 -- 30 $M_{\odot}$ will evolve to blue supergiants (BSGs) of spectral types B and A, before cooling further and becoming red supergiants (RSGs) \citep[e.g.][]{1981A&A...102..401M, 1986ARA&A..24..329C, 2000ARA&A..38..143M, 2008ApJ...684..118U, 2008MNRAS.384.1109E,  2012A&A...537A.146E, 2012ARA&A..50..107L, 2013ApJ...764...21C}.
While BSGs have classically been considered post core-H burning stars, recent studies have raised the possibility that BSGs are actually still core-H burning \citep{2015A&A...575A..70M, 2010A&A...512L...7V}.
With temperatures ranging from 8\,000~K to 25\,000~K, the spectrum of the B and A type BSGs peaks at visual wavelengths. They are among the brightest stars in the universe in the visual range, with absolute visual magnitudes of up to $M_{\rm V} = -9.5$ mag.

Observational evidence for a large number of binaries has been around for many decades \citep[e.g.][]{1980ApJ...242.1063G}, but recently our understanding of the importance of binaries has rapidly increased \citep{2007ApJ...670..747K, 2009AJ....137.3358M, 2012Sci...337..444S, 2014ApJS..211...10S, 2015A&A...580A..93D, 2017ApJS..230...15M, 2017A&A...598A..84A}. Interaction between stars in binary systems can lead to mass transfer episodes, typically classified into Case A (pre-core hydrogen exhaustion), Case B (post-core hydrogen exhaustion) or Case C (post-core He burning) mass transfer \citep{1966AcA....16..231P, 1967ZA.....65..251K}. Stellar evolution models and population synthesis of massive stars have also concluded that some observations are better explained by incorporating binaries \citep[e.g.][]{1998A&ARv...9...63V, 1998A&A...334...21V, 2005A&A...435.1013P, 2008MNRAS.384.1109E, 2013ApJ...764..166D, 2017A&A...601A..29Z, 2017ApJ...840...10Y}.

Recent observational campaigns have improved our understanding of initial parameters of binary systems, in particular the fraction of stars that exist in a binary system. Observations by \citet{2012Sci...337..444S} reported an interacting binary fraction of 0.69 $\pm$ 0.09 for O-type stars in open clusters in the Galaxy. \citet{2013A&A...550A.107S} and \citet{2015A&A...580A..93D} studied the Tarantula region in the LMC, reporting interacting binary fractions of 0.51 $\pm$ 0.04 for O-type stars and 0.58 $\pm$ 0.11 for B-type stars respectively.
%\citet{2013A&A...550A.107S} reported an interacting binary fraction of 0.51 $\pm$ 0.04 for the Tarantula region in the LMC and \citet{2015A&A...580A..93D} reported a binary fraction of 0.58 $\pm$ 0.11 for B-type stars.} 
%\citet{2013A&A...550A.107S} and \citet{2015A&A...580A..93D} studied the Tarantula region in the LMC, reporting interacting binary fractions of 0.51 $\pm$ 0.04 for O-type stars and 0.58 $\pm$ 0.11 for B-type stars respectively.
\citet{2017ApJS..230...15M} compiled observations of early-type binaries and reported a single star fraction of 16\% for 9 -- 16 $M_{\odot}$ stars and 6\% for > 16 $M_{\odot}$ stars (these values include non-interacting long period binaries). Because the fraction of massive stars that exist in binary systems is high, understanding the effects of binary interactions on both the primary and the secondary is critical to understanding the evolution of massive stars and their impact on host galaxies.

Because BSGs are typically the brightest stars in their galaxies in optical light, they are ideal candidates for determining extragalactic distances \citep{1999A&A...350..970K}. The flux-weighted gravity luminosity relationship (FGLR) is a powerful method which has been used to determine extragalactic distances with BSGs, both within the Local Group \citep{2008ApJ...684..118U, 2009ApJ...704.1120U} and beyond \citep{2012ApJ...747...15K, 2014ApJ...788...56K, 2016ApJ...829...70K, 2016ApJ...830...64B}.
The FGLR is an observationally tight relationship for BSGs, between the absolute bolometric magnitude $M_{\rm bol}$ and the "flux-weighted gravity" $g_{\rm F}$. For a star with a given surface gravity g, and effective temperature $T_{\rm eff}$, \citet{2003ApJ...582L..83K} defined

\begin{equation}
g_{\rm F} = g/T^4_{\rm eff}.
\end{equation}

The technique was first described by \citet{2003ApJ...582L..83K} and has since been further calibrated \citep{2008ApJ...681..269K, 2017AJ....154..102U}.
A major advantage of the FGLR technique for distance measurement is that it uses a spectroscopic method that allows for accurate correction of interstellar reddening and extinction of each individual blue supergiant. This avoids extinction and metallicity induced uncertainties associated with alternative distance methods.

It has been shown that the shape, scatter and metallicity dependence of the observed FGLR are well explained by stellar evolution models of single stars \citep{2015A&A...581A..36M}.
However, some evolutionary scenarios such as those with very strong mass loss during the RSG stage (causing the star to evolve back to the blue with a low mass) were found to be disfavoured. 
If frequent, such scenarios would produce a scatter of the FGLR well above the one that is observed \citep{2015A&A...581A..36M}. The FGLR was also briefly studied using evolutionary models of binary systems by \citet{2017PASA...34...58E}.  
%Given that we expect a large fraction of the observed BSGs to exist in binary systems, it is worthwhile to investigate the effects of binary interaction on the evolution of BSGs.

In this paper, we expand upon the work of \citet{2015A&A...581A..36M} and investigate the impact of close binary evolution on the FGLR.
In particular, we explore the question of whether Roche-lobe overflow produces BSGs that are compatible with the observed FGLR.
%In particular, mass transfer episodes in binary systems due to Roche-Lobe overflow may produce BSGs that are incompatible with the observed FGLR.
%In particular, close binary evolution may produce strong mass loss during the RSG stage as a result of mass transfer and thus might produce BSGs that are incompatible with the observed FGLR. 
We use the detailed library of tracks in v2.1 BPASS computed by \citet{2017PASA...34...58E}, and our own close binary stellar evolution models that we compute using MESA \citep{2015ApJS..220...15P,2013ApJS..208....4P,2011ApJS..192....3P}.

The paper is organised as follows. In Sect. 2 we describe the properties of the stellar evolution models. In Sect. \ref{sec:binint}, we investigate how the evolution of BSGs is affected by binary interaction. We discuss the results and implications in Sect. \ref{sec:discussion} and conclude our analysis in Sect. \ref{sec:conclusions}.

%__________________________________________________________________

\section{Stellar evolution models} \label{sec:models}
We use the BPASS suite of binary models and the MESA stellar evolution code to study the properties of BSGs in binary systems. To reproduce the bulk of the observed population, we focus on a mass range of 9 -- 30 $M_{\odot}$. We study models with initial orbital periods in the range log($P$/days) = 1.0 -- 3.4 (i.e. P = 10 -- 2511 days).
As we discuss in Sect. \ref{sec:massratio}, the mass ratio $q$ (where $q = m_{\rm sec}/m_{\rm pri}$) has little effect on the evolution of the primary star in the BSG stage and on the FGLR plot over a wide range of initial masses and periods. For this reason, we discuss only a mass ratio of $q$ = 0.9 throughout this paper.

\subsection{BPASS models}
We select models with initial primary masses of 9, 15, 20 and 30 $M_{\odot}$, metallicity Z = 0.020 and mass ratio $q$ = 0.9 from the v2.1 BPASS suite of binary models \citep[see][for details]{2017PASA...34...58E}.
To examine the effects of the initial period, we choose models with initial orbital periods of log($P$/days) = 1.4, 2.0, 2.4, 3.0, 3.2 and 3.4.

%We use the v2.1 BPASS suite of binary models \citep[see][for details]{2017PASA...34...58E} with initial primary masses of 9, 15, 20 and 30 $M_{\odot}$ and with metallicity Z = 0.020.
%To examine the effects of the initial period of the binary system, we choose initial orbital periods of log($P$/days) = 1.4, 2.0, 2.4, 3.0, 3.2, 3.4.
%As we will discuss in Sect. \ref{sec:massratio}, the mass ratio $q$ (where $q$ = $m_2/m_1$, for a primary and secondary mass of m$_1$ and m$_2$ respectively) has little effect on the evolution of the primary star in the BSG stage and on the FGLR plot over a wide range of initial masses and periods. For this reason, we will discuss only a mass ratio, $q$ = 0.9  in the main body of this paper.

\subsection{MESA models}
To complement the BPASS models, we use the MESA stellar evolution code to compute our own models \citep{2015ApJS..220...15P,2013ApJS..208....4P,2011ApJS..192....3P}.
The models in the BPASS suite follow only the primary star with detailed calculations \citep{2017PASA...34...58E}. It is important to investigate whether the secondary will produce a BSG and how these BSGs compare with the observed data. For this reason, we compute a small grid of models using MESA. Our models have initial primary masses of 12, 15, 20 and 30 $M_{\odot}$ with metallicity Z = 0.020 and a mass ratio of $q$ = 0.9. 
We choose a range of initial orbital periods of log($P$/days) = 1.0, 2.0, 2.6, 3.0, 3.1, 3.3 and 3.4.
We exclude very short periods as the system is likely to enter common envelope evolution, which is difficult to model accurately.
We also exclude systems with initial orbital periods $P >$ 2511 days, because the stars in these systems would evolve very similarly to single stars.

\subsection{Physical ingredients of models}\label{sec:modingredients}
In this section, we summarise the physical ingredients we use in our MESA models and provide a comparison with the ingredients in the BPASS models.
\begin{itemize}
\item[--] We use the Schwarzschild criterion for convection (as in BPASS).\\
\item[--] We consider convective core overshoot using a step function over a layer of thickness 0.3 $H_{\rm P}$ above the hydrogen core, where $H_{\rm P}$ is the pressure scale height at the outer boundary of the core. The BPASS models include convective overshooting with $\delta_{\rm ov}$ = 0.12, which results in an overshooting length of around 0.3 $H_{\rm P}$ for massive stars.\\
\item[--] We use a mixing length for convection of 1.5 $H_{\rm P}$. The BPASS models use a mixing length for convection of 2.0 $H_{\rm P}$. We use the MLT++ scheme \citep{2013ApJS..208....4P} in MESA to assist with the convergence of the models. We note that this may impact the properties of BSGs after mass transfer.\\
\item[--] We use the `Dutch' wind mass loss scheme in MESA with the scaling factor of 1.0. This scheme involves a combination of results from \citet{1988A&AS...72..259D, 2000A&A...360..227N, 2001A&A...369..574V} for different regimes. A similar mass loss scheme is used in BPASS.\\
\item[--] We use the `Kolb' \citep{1990A&A...236..385K} mass transfer prescription to calculate the mass transfer rate. The BPASS models use a method inspired by \citet{2002MNRAS.329..897H} with a Roche lobe radius defined by \citet{1983ApJ...268..368E}.\\
\item[--] We do not consider rotation or tidal interactions, but assume non-conservative mass transfer as in \citet{2017ApJ...840...10Y} and as predicted by previous binary models including the effects of rotation \citep{2010ApJ...725..940Y, 2005A&A...435.1013P}. In these studies, the accretor is quickly spun up to critical rotation during mass transfer and in response, the stellar wind mass loss increases dramatically. This effectively results in highly non-conservative mass transfer. Our models all undergo Case B mass transfer and we use a mass accretion efficiency of $\beta = 0.2$ as suggested by previous results \citep[e.g.][]{2010ApJ...725..940Y}. The BPASS models assume a maximum accretion rate $\dot{M}_{\rm sec, \, max} = M_{\rm sec}/\tau_{\rm KH}$, where $\tau_{\rm KH}$ is the Kelvin-Helmholz timescale.\\
\item[--] We evolve the stars to a central temperature of $T_{\rm c} = 10^9$~K, corresponding to the end of carbon burning. The luminosity and effective temperature remain almost constant after $T_{\rm c}$ increases beyond $10^9$~K \citep{2004A&A...425..649H, 2013A&A...558A.131G, 2017ApJ...840...10Y}.
\end{itemize}

In this work, we define the BSG stage for the primary stars as post-main sequence stars with an effective temperature between 8\,000 and 25\,000~K, and hydrogen surface fraction $X$ > 0.5. 
Due to mass accretion, the secondary stars may expand and look like BSGs before the completion of core hydrogen burning. For this reason, we define the secondary stars with $T_{\rm eff}$ between 8\,000 and 25\,000~K, and $X$ > 0.5, as BSGs if they have accreted mass due to mass transfer, even if they are still core-hydrogen burning stars. 
%We define the BSG stages for the secondary using the same conditions but, as the secondary stars possibly form BSGs before completion of core hydrogen burning, it is unclear if and when they will be blue supergiants. Therefore, we define the secondary stars as BSGs when they leave the typical single star main sequence (MS) track due to mass accretion.

\begin{figure}
	\centering
	\includegraphics[width=\hsize]{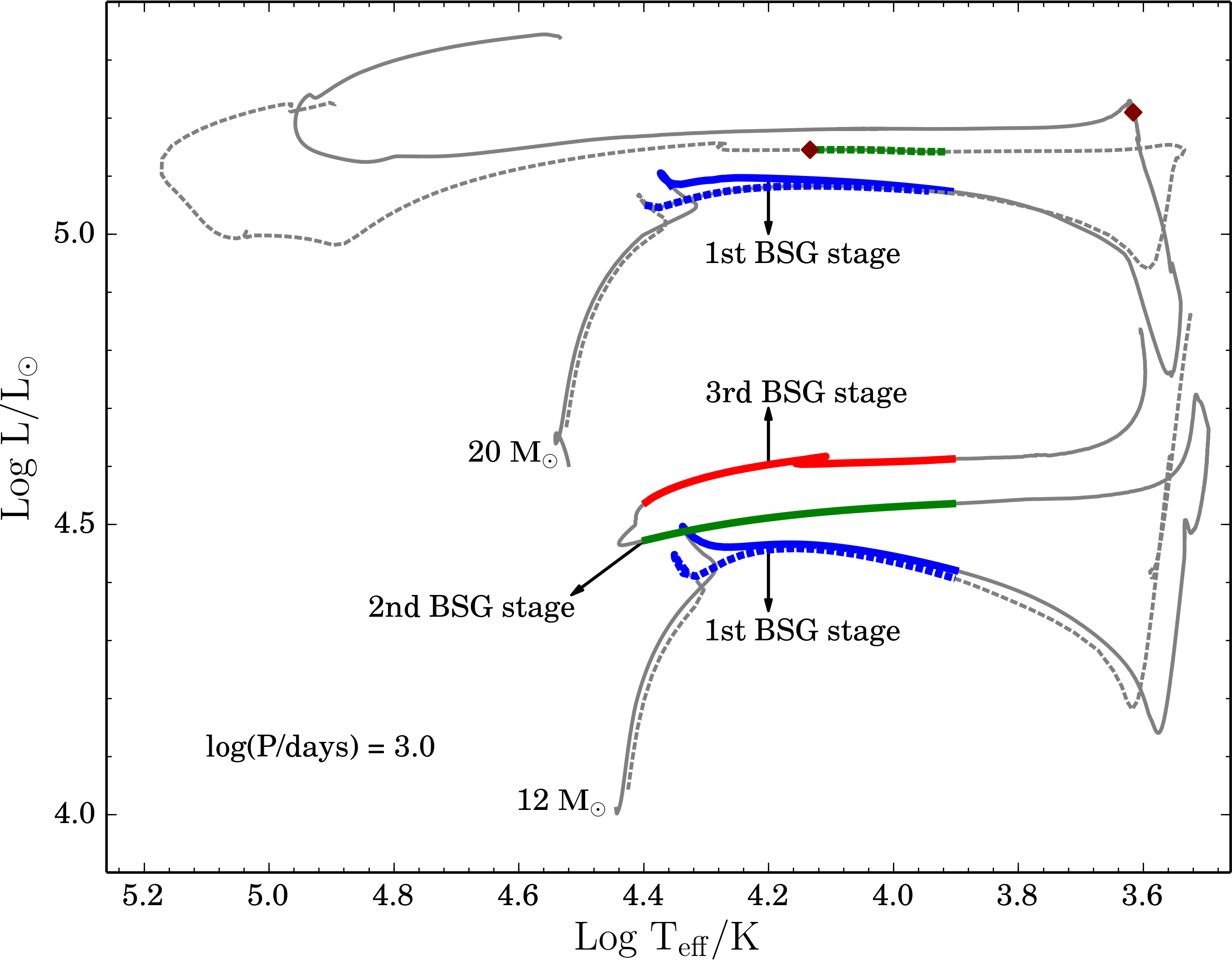}
	\caption{Evolutionary tracks for primary stars of masses 12 and 20 $M_{\odot}$ for BPASS models (dashed line) and MESA models (solid line). The initial orbital period is log($P$/days) = 3.0. Blue, green and red indicate the first, second and third BSG stages respectively. The maroon diamond indicates the evolutionary point at which the hydrogen surface abundance drops below 0.5.}
	\label{fig:comp}
\end{figure}

In Fig. \ref{fig:comp} we compare the BPASS and MESA evolutionary tracks for models with the same initial mass of the primary, mass ratio and initial orbital period. For the 20 $M_\odot$ models, BPASS and MESA produce similar qualitative evolution, however there are some quantitative differences between the models due to differences in physical ingredients. The main sequence (MS) track in the MESA model is slightly longer than the BPASS model because the MESA models were computed with a different overshoot implementation, creating a larger convective core during the core-H burning phase. At this point it is worth noting that different stellar evolution codes use different values for the overshooting parameter as well as different implementations of convective-core overshooting. This has consequences for the MS lifetime and the width of the MS \citep{2013A&A...560A..16M}. Increased overshooting results in larger cores and an extension of the MS to cooler temperatures. The extension of the MS also depends on other complex mechanisms such as rotation and magnetic fields.

The 20 $M_\odot$ MESA model undergoes a sharp drop in luminosity before the RSG stage due to a mass transfer episode. Such a sharp drop in luminosity is not present in the BPASS models. This is likely due to differences in mass transfer prescriptions adopted in the MESA models and the BPASS models. The evolutionary point at which $X$ drops below 0.5 occurs during the RSG stage in the MESA model and in the blue part of the HR diagram in the BPASS model. This difference is also likely largely due to overshooting. Larger cores (as in the MESA models) make the star evolve more rapidly to the red part of the HR diagram after the MS phase. This means that a larger fraction of the core helium burning phase occurs during the RSG phase where strong mass losses occur. These strong mass losses favour a more rapid appearance of deep layers at the surface. We also note that the BPASS model produces a hotter Wolf-Rayet (WR) star than the MESA model. A possible explanation is that the two models have different surface compositions due to different mass-loss histories.

The 12 $M_\odot$ BPASS and MESA models differ qualitatively in their post-MS evolution.
The MESA model produces a well developed loop but the BPASS model remains in the RSG phase. 
This is a result of the fact that the MESA models are redder and more extended than the BPASS models (due to differences in mixing lengths for convection) and therefore more prone to go through a stronger mass transfer episode during the RSG phase.

\section{Impacts of binary interaction on blue supergiants}\label{sec:binint}
In this section we discuss the impact of binary interaction on BSGs under the assumption that the BSG can be observed without spectral contamination from its companion. In Sect. \ref{sec:unresol_dis}, we discuss the impact of the presence of a secondary on the quantities inferred from a combined unresolved spectrum.

\subsection{Primary stars}

The left panels of Figs. \ref{fig:bpassmodels} and \ref{fig:mesaprimmodels} show the evolutionary tracks of the primary stars in the HR diagram from the BPASS models and the MESA models respectively. A star can exist as a BSG either when crossing the HR diagram from the main sequence to the RSG stage (first stage), or after the RSG stage (second stage).

\subsubsection{First stage BSGs}
With both the BPASS and MESA models, BSGs that form during the first crossing of the HR diagram after the main sequence (first stage) produce characteristics mostly consistent with the observed FGLR (right panels of Figs. \ref{fig:bpassmodels} and \ref{fig:mesaprimmodels}). In systems with initial periods $P >$ 100 days, the first BSG stage is not affected by binarity as the mass transfer does not begin until after the BSG stage, when log($T_{\rm eff}$/K) $<$ 3.9. In these cases, these blue supergiants present similar characteristics to those obtained from a single star.

However, in systems with short initial periods, P $\leq$ 100 days, the mass transfer begins during or before the first BSG stage (e.g. see dashed blue part of tracks in Fig. \ref{fig:mesaprimmodels}). This is evident in the HR diagram from the drop in luminosity as a result of the mass loss. In the BPASS models, despite the change in mass (which results in a change in $g_{\rm F}$), the BSGs are still compatible with the observed FGLR due to a corresponding change in luminosity. For the MESA models, this large mass transfer produces tracks outside the observed FGLR.  However, because Roche-lobe overflow (RLOF) is taking place, the spectrum of the star would probably be modified because of the high accretion and mass-loss rate (see discussion for post-mass transfer systems in Sect. \ref{sec:further}). 
As a result, these stars would probably not appear as a normal BSG and therefore the FGLR would not apply to them. In this case, the duration of the BSG stage would be shortened because of the beginning of the mass transfer episode.

\subsubsection{Second stage BSGs} \label{sec:secondstagebsg}
%A second BSG stage may be formed by stars that evolve back to the blue region of the HR diagram after the RSG phase.
Our models show that not all evolutionary tracks going back to the blue after a RSG stage produce a BSG phase. This is because some models have low hydrogen surface abundances, X, and we only consider stars with $X$ $>$ 0.5 as BSGs.

If we compare the post-RSG evolution of the 15 $M_{\odot}$ models computed with MESA for different initial periods (see left panel of Fig. \ref{fig:mesaprimmodels}), we find that there are three different evolutionary outcomes. For a system with a short initial period (e.g. log($P$/days) = 2.0), the star evolves back to the blue after the RSG stage, but with $X$ < 0.5 so we do not classify it as a BSG. These stars with hydrogen-poor envelopes would likely be classified as blue hypergiants or luminous blue variables. For a system with an intermediate period (e.g. log($P$/days) = 3.0), the star evolves back towards the blue with $X$ $\geqslant$ 0.5 creating a second BSG stage. For a system with a long period (e.g. log($P$/days) = 3.4), the star does not evolve back to the blue. This trend is due to a general decrease in mass loss due to mass transfer, with increasing initial orbital period.

\begin{figure*}[!htbp]
	\centering
	\includegraphics[width=.45\textwidth]{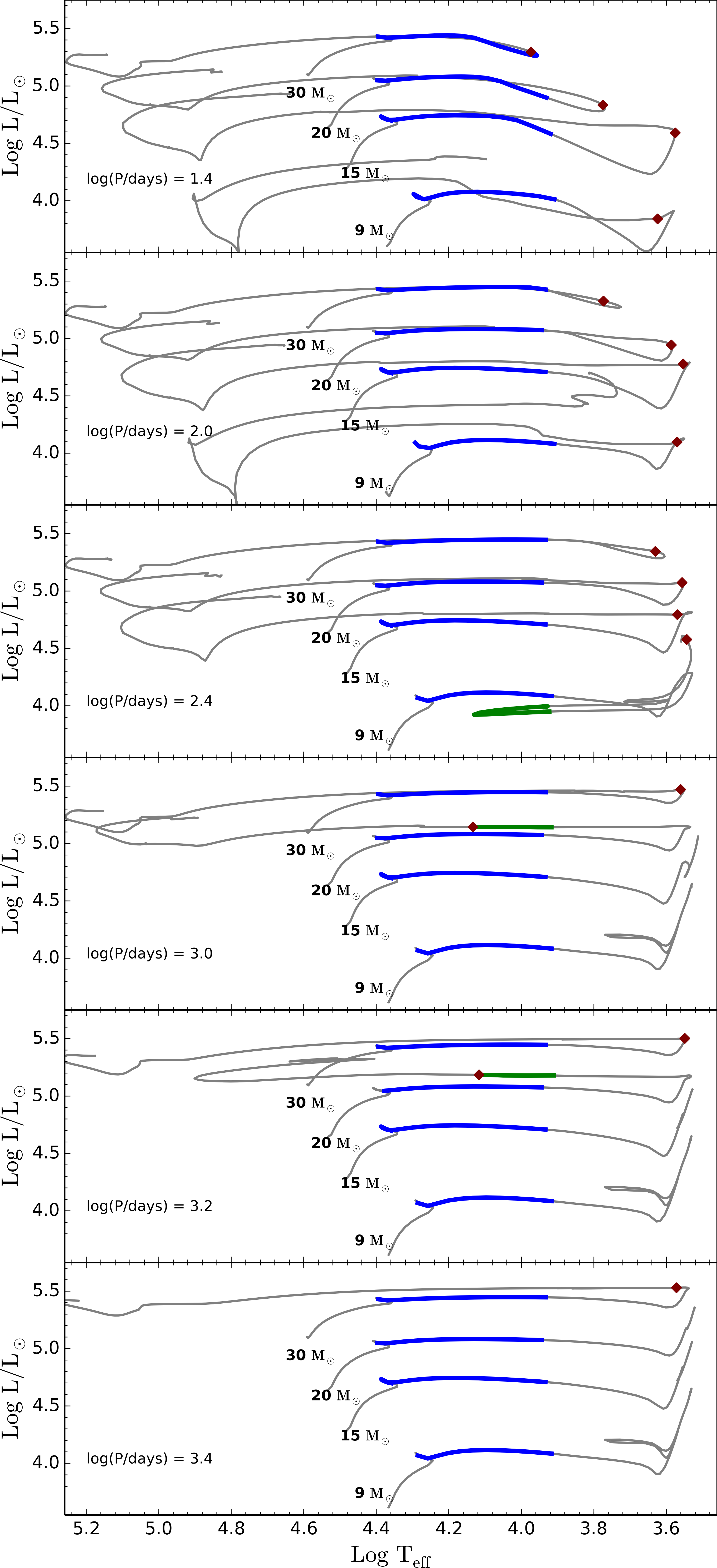} \includegraphics[width=.467\textwidth]{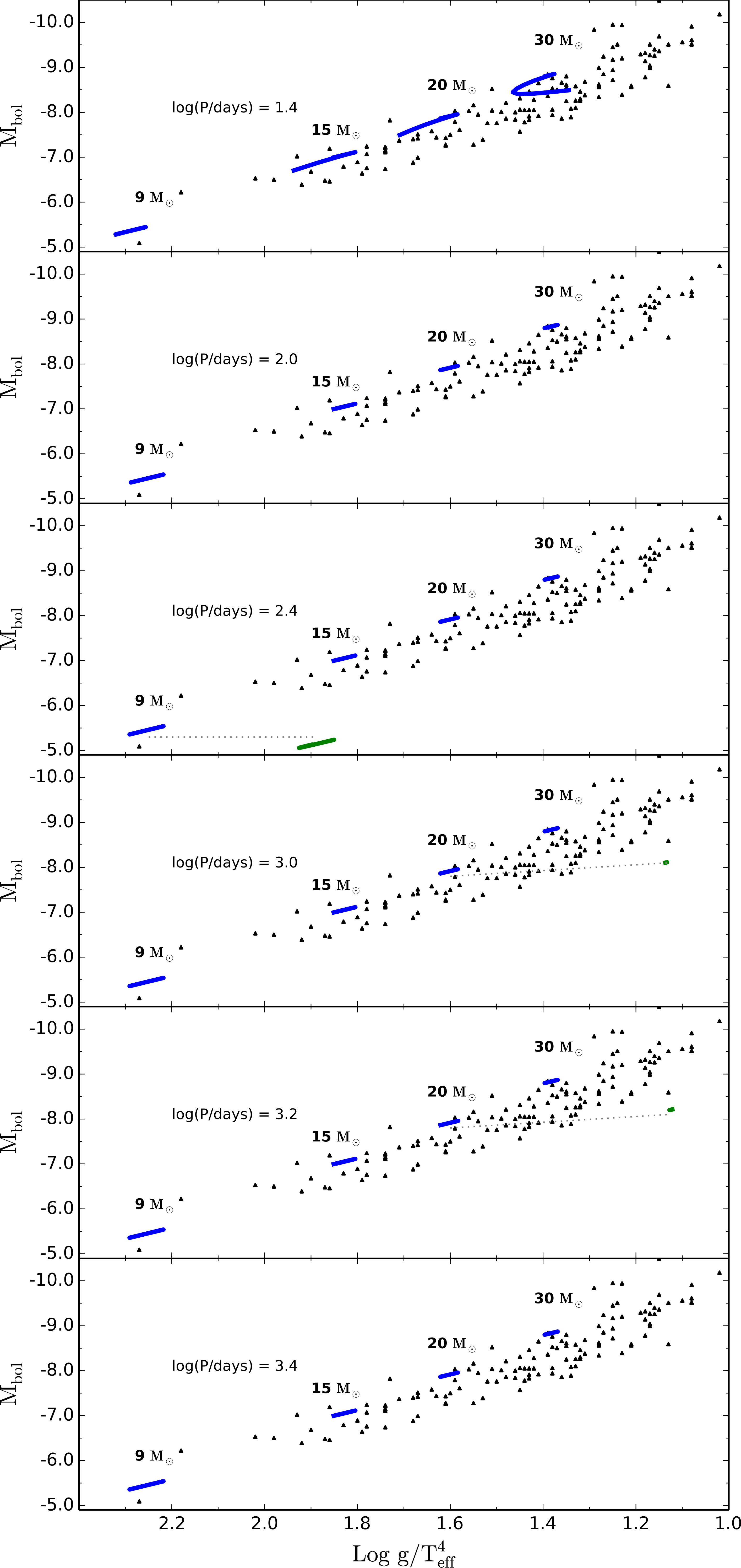}
	\caption{{\it Left panel:} HR diagram for BPASS models (primary stars only) with initial periods of log($P$/days) = 1.4, 2.0, 2.4, 3.0, 3.2, 3.4, where P is in days. Initial masses are 9, 15, 20 and 30 $M_{\odot}$ with a mass ratio of 0.9. Blue and green indicate first and second BSG stages respectively. Not all parts of the track crossing the BSG after a RSG stage are in green. This is because we define that a BSG has a hydrogen surface fraction $X$ > 0.5. The maroon diamond indicates the evolutionary point at which the hydrogen surface abundance drops below 0.5. {\it Right panel:}  $M_{\rm bol}$ vs. Log $g/T_{\rm eff}^4$ planes for the same models as in the left panel. The colours have the same meaning as the left panel.
Black triangles represent observations of individual BSGs. The observations are taken from NGC 300 \citep{2008ApJ...681..269K}, other galaxies \citep{2008PhST..133a4039K}, M33 \citep{2009ApJ...704.1120U}, M81 \citep{2012ApJ...747...15K}, WLM \citep{2008ApJ...684..118U}, NGC 3109 \citep{2014ApJ...785..151H}, NGC 3621 \citep{2014ApJ...788...56K} and NGC 4258 \citep{2013ApJ...779L..20K}. Light grey dashed lines join the first and second BSG stages from the same model.}
	\vspace{80pt}
	\label{fig:bpassmodels}
\end{figure*}

\begin{figure*}[!htbp]
	\centering
	\includegraphics[width=.48\textwidth]{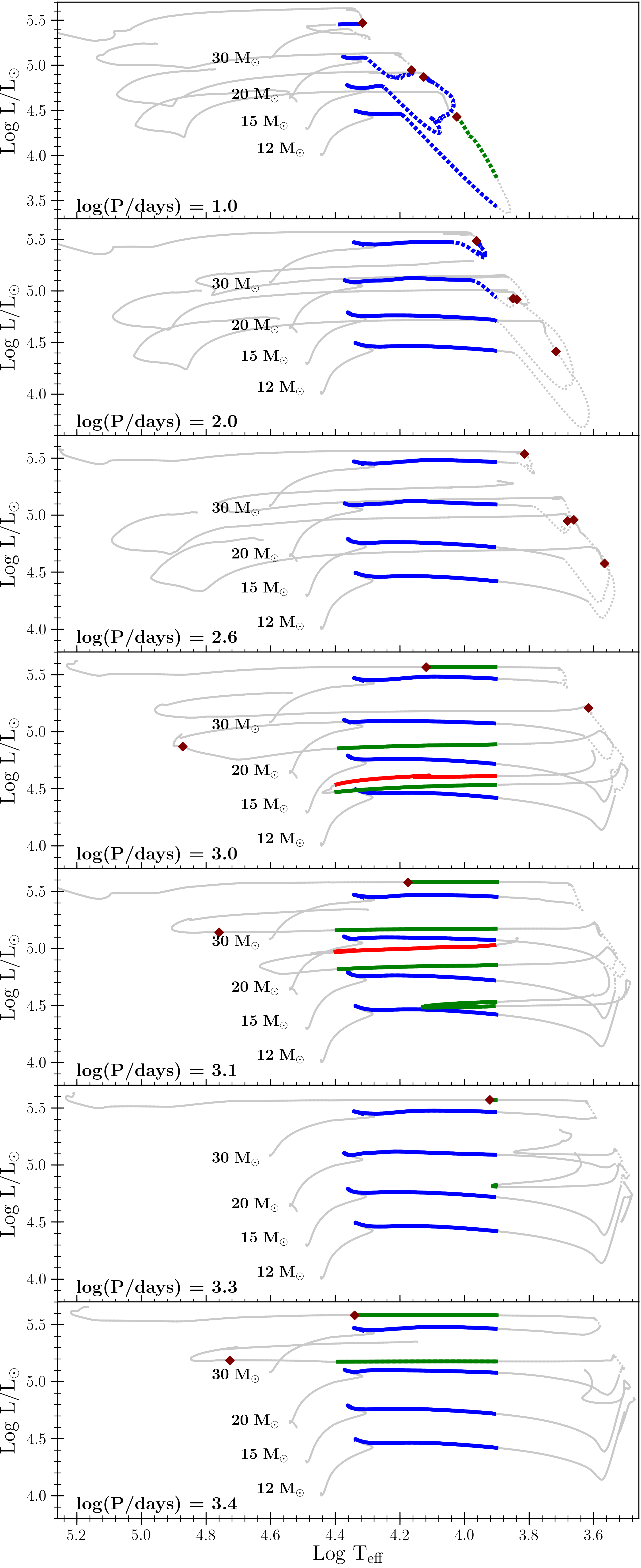} \includegraphics[width=.486\textwidth]{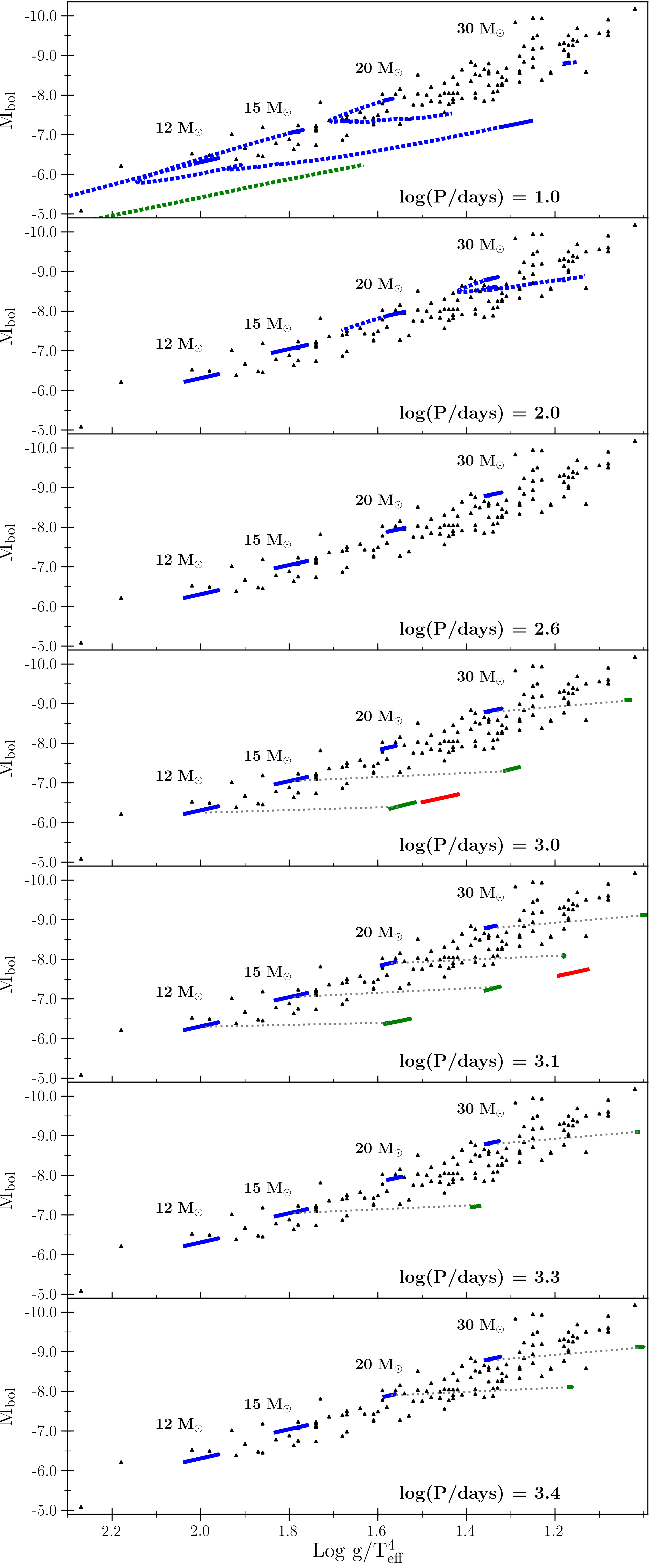}
	\caption{{\it Left panel:} HR diagrams for primary stars in MESA models, with initial masses of 12, 15, 20 and 30 $M_{\odot}$, mass ratio of $q$ = 0.9 and Z = 0.020. Blue, green and red indicate first, second and third BSG stages respectively. The dashed line indicates the period during which mass transfer takes place. Not all parts of the track crossing the BSG after a RSG stage are in green. This is because we define that a BSG has a hydrogen surface fraction $X$ > 0.5. The maroon diamond indicates the evolutionary point at which the hydrogen surface abundance drops below 0.5. {\it Right panel:} $M_{\rm bol}$ vs. Log $g/T_{\rm eff}^4$ planes for same models as in left panel. The colours have the same meaning as the left panel. Observations sources are listed in caption of Fig. \ref{fig:bpassmodels}. Light grey dashed lines join the first and second BSG stages from the same model.}
	%\vspace{10pt}
	\label{fig:mesaprimmodels}
\end{figure*}

Interestingly, BSGs that result from a post-RSG stage evolution are produced with characteristics far away from the observed FGLR (see the green parts of the tracks in the FGLR planes in Figs. \ref{fig:bpassmodels} and \ref{fig:mesaprimmodels}). In the BPASS models, these BSGs occur for only a relatively narrow range of parameters; only the 9 and 20 $M_{\odot}$ models produce a second BSG stage and these are only produced for a limited range of periods. More second stage BSGs are produced in the MESA models. This is due to different ingredients in the physical models such as mixing and mass transfer. These second stage BSGs are produced after undergoing strong mass loss during the RSG stage. Recalling that  $g_{\rm F}$ = $g/T^4_{\rm eff} \propto M/L$, we see that the combination of the decreased mass and slightly increased luminosity results in a lower flux-weighted gravity, $g_{\rm F}$. Because $g_{\rm F}$ decreases and the luminosity increases only slightly, the track in the FGLR plane is shifted to the right.
With the BPASS models, \citet{2017PASA...34...58E} reproduced the FGLR for primary stars using a population synthesis. In general they found that their models were consistent with the observed FGLR, but they also noted the presence of stars to the right of the observed scatter in the FGLR plane. This is consistent with the results obtained here.

Whether a primary star of a given mass will produce a second BSG stage depends on the initial period of the system.
Using this, we can estimate the percentage of primary stars, for a given mass, that are positioned away from the observed FGLR. We can combine the initial period distribution for binary systems, the range of periods that produce a second BSG stage given by the models (Figs. \ref{fig:bpassmodels} and \ref{fig:mesaprimmodels}) and the lifetimes of these stages.
%Using this, we can combine the initial period distribution for binary systems with the range of periods that produce a second BSG stage given by the models (Figs. \ref{fig:bpassmodels} and \ref{fig:mesaprimmodels}) and the lifetimes of these stages, to estimate the percentage of primary stars, for a given mass, that are positioned away from the observed FGLR.
We use the initial period distribution for binary systems reported by \citet{2012Sci...337..444S}, $f(\mathrm{log}(P/\mathrm{days})) \propto \mathrm{log}(P)^{-0.55}$ for log($P$/days) $\in$ [0.15, 3.5]. 
Based on the MESA models, we approximate that 30 $M_{\odot}$ primary stars will produce a second BSG stage for 2.8 < log($P$/days) < 3.5.
The lifetime of this second BSG stage as a percentage of the total BSG lifetime of the star is typically 80\% in the MESA models. 
Given the above period range, period distribution and lifetime, and assuming all 30 $M_{\odot}$ stars are primary stars that exist in binary systems with log($P$/days) $\in$ [0.15, 3.5], we expect 10\% of 30 $M_{\odot}$ primary stars to be located away from the tight scatter in the FGLR plane. This percentage is an upper limit, as not all stars are in a binary system \citep{2017ApJS..230...15M, 2012Sci...337..444S}. This percentage is similar for other primary masses in the MESA models investigated in this paper.
Applying the same method to the BPASS models, we estimate that 4\% of 9 $M_{\odot}$ and 1\% of 20 $M_{\odot}$ primary stars should be located away from the tight scatter in the FGLR plane. Therefore, based on the stellar evolution models and the above assumptions, we would expect to see some stars located away from the observed FGLR in the sample of \textasciitilde 140 BSGs shown in the $M_{\rm bol}$ vs. log $g/T^4_{\rm eff}$ diagrams in Figs. \ref{fig:bpassmodels} and \ref{fig:mesaprimmodels}. A few such objects have been detected and proposed to be products of binary interaction \citep{2009ApJ...704.1120U}, however they are not included in the sample of BSGs shown in Figs. \ref{fig:bpassmodels}, \ref{fig:mesaprimmodels}, \ref{fig:mesasecmodels}. We discuss this further in Sect. \ref{sec:discussion}.

\subsection{Secondary stars}
\begin{figure*}[!htbp]
	\centering
	\includegraphics[width=.48\textwidth]{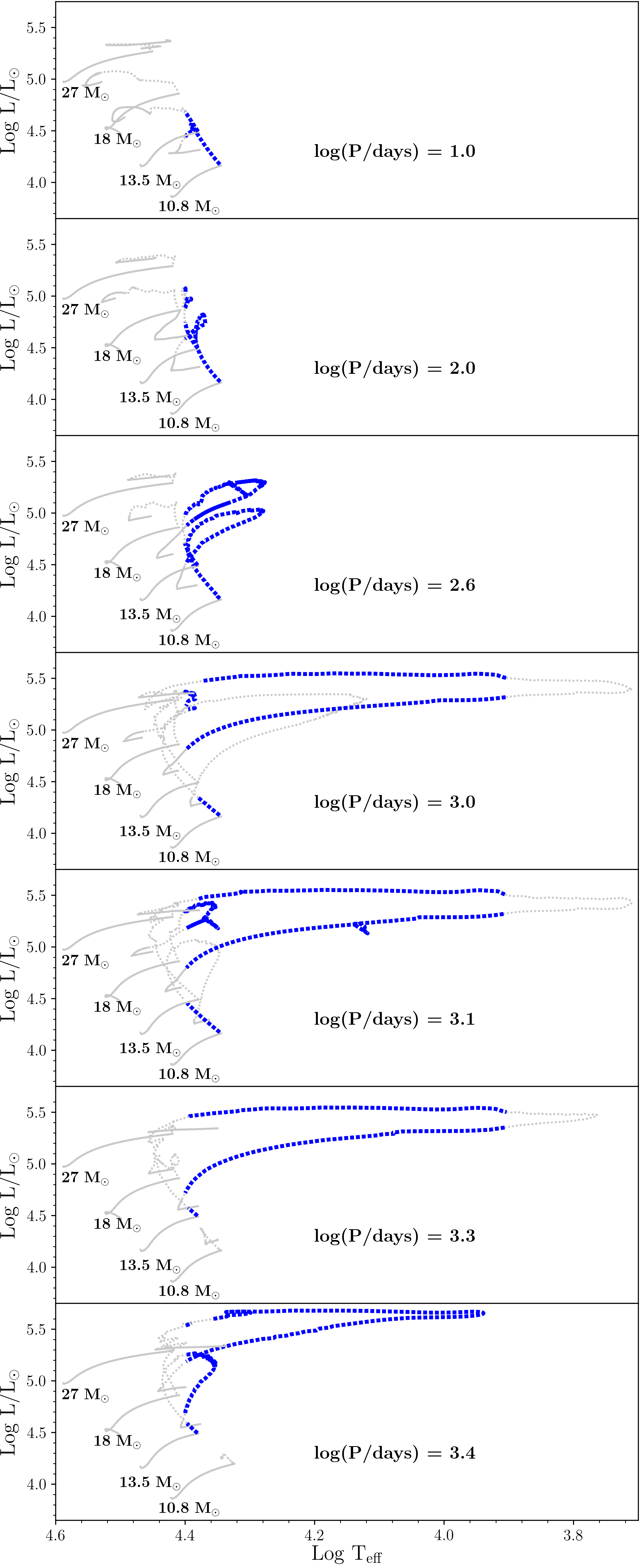} \includegraphics[width=.486\textwidth]{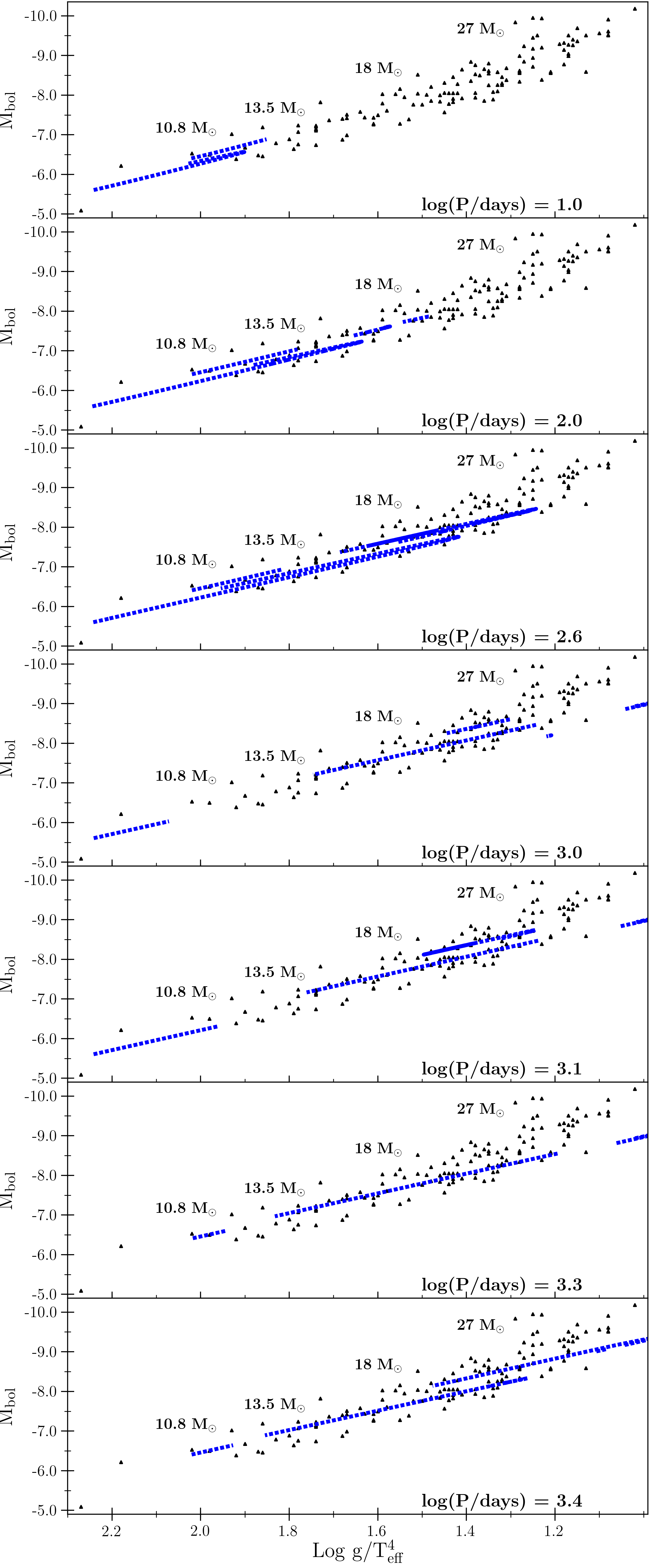}
	\caption{{\it Left panel:} HR diagrams for secondary stars in MESA models, with initial masses of 10.8, 13.5, 18 and 27 $M_{\odot}$. See caption of Fig. \ref{fig:mesaprimmodels} for other details. {\it Right panel:} $M_{\rm bol}$ vs. Log $g/T_{\rm eff}^4$ planes for same models as in left panel. Observations sources are listed in caption of Fig. \ref{fig:bpassmodels}.}
	\vspace{61pt}
	\label{fig:mesasecmodels}
\end{figure*}

The left panel of Fig. \ref{fig:mesasecmodels} shows the evolution of the secondary stars during the lifetime of the primary. The secondary stars begins its evolution on the MS, the same as single stars. When mass transfer begins due to RLOF from the primary, the luminosity and temperature increase. For systems with shorter periods, the secondary stars stay in the blue part of the HR diagram. For some systems with larger periods, the secondary stars may expand rapidly towards the red and then contract back towards the blue as they readjust to the increased mass (e.g. 13.5 $M_{\odot}$ with log($P$/days) = 3.3). The consequences of mass accretion for the evolution of the secondary stars depend on the amount of mass transferred, the structure of the envelope when mass transfer takes place and the mass transfer mechanism used in the evolution code (see Sect. \ref{sec:modingredients}). For classical studies of accretion onto secondary stars, we refer the reader to \citet{1976ApJ...206..509U, 1977A&A....54..539K}. We also note the topic has been more recently discussed by \citet{2007A&A...467.1181D, 2007A&A...465L..29C}.

The predicted FGLR sequences for these secondary stars (assuming the BSG definition given above) are in good agreement with the observed data. This agreement with the observations is helped by the low mass accretion efficiency (0.2) that we use in our MESA models. As indicated by the dashed lines in Fig. \ref{fig:mesasecmodels}, almost all of the BSG stages occur during a regime where the secondary is technically (according to our criteria) a BSG, but it is interacting with the primary. During this interaction, the models indicate that, for a short period of time, the mass accretion rates can reach $10^{-2} M_{\odot}$/yr which may obscure the star.
%As the mass accretion rates can reach $\mathrm{10^{-2}M_{\odot}/yr}$, this interaction may either obscure the star meaning it would not be considered in the FGLR.
The secondary may be spun up to the critical rotation as the transferred matter carries angular momentum. The high mass accretion rate may also produce a dense, flattened keplerian disk around the secondary, as in the case of HD327083 \citep{2012A&A...538A...6W, 2012A&A...543A..77W}. Furthermore, some B[e] stars have been resolved in binaries \citep{2017ASPC..508..163M}. Detailed radiative transfer models would be needed to test this scenario.
%The star could potentially appear as a B[e] star because it is spun up to the critical rotation as the transferred matter carries angular momentum. The high mass accretion rate may also produce a dense, flattened circumstellar medium around the secondary, where H and forbidden lines that are characteristic of B[e] stars would be formed. Detailed radiative transfer models are needed to test this scenario.

When the primary star dies, a system composed of a neutron star and the secondary star is formed. We continue to follow the evolution of the binary system with MESA (treating the neutron star as a point mass) until the secondary reaches the end of carbon burning. We estimate the mass of the neutron star using the final CO core mass of the primary and the relationship between the remnant mass and the CO core mass given in Table 4 in \citet{2012A&A...542A..29G}. We calculate the initial separation for the post-supernova evolution using the final separation from the pre-supernova models and taking the envelope mass ejected by the primary during the neutron star formation into account. We assume circular orbits and no neutron star kick due to the supernova. A significant fraction of such binary systems would be unbound because of the neutron star kick, and the secondary would become a single star. For detailed studies of the consequences of neutron star kicks and non-circular orbits see, for example, \citet{1998A&A...330.1047T} and \citet{2018arXiv180409164R}.

\begin{figure}
	\centering
	\includegraphics[width=\hsize]{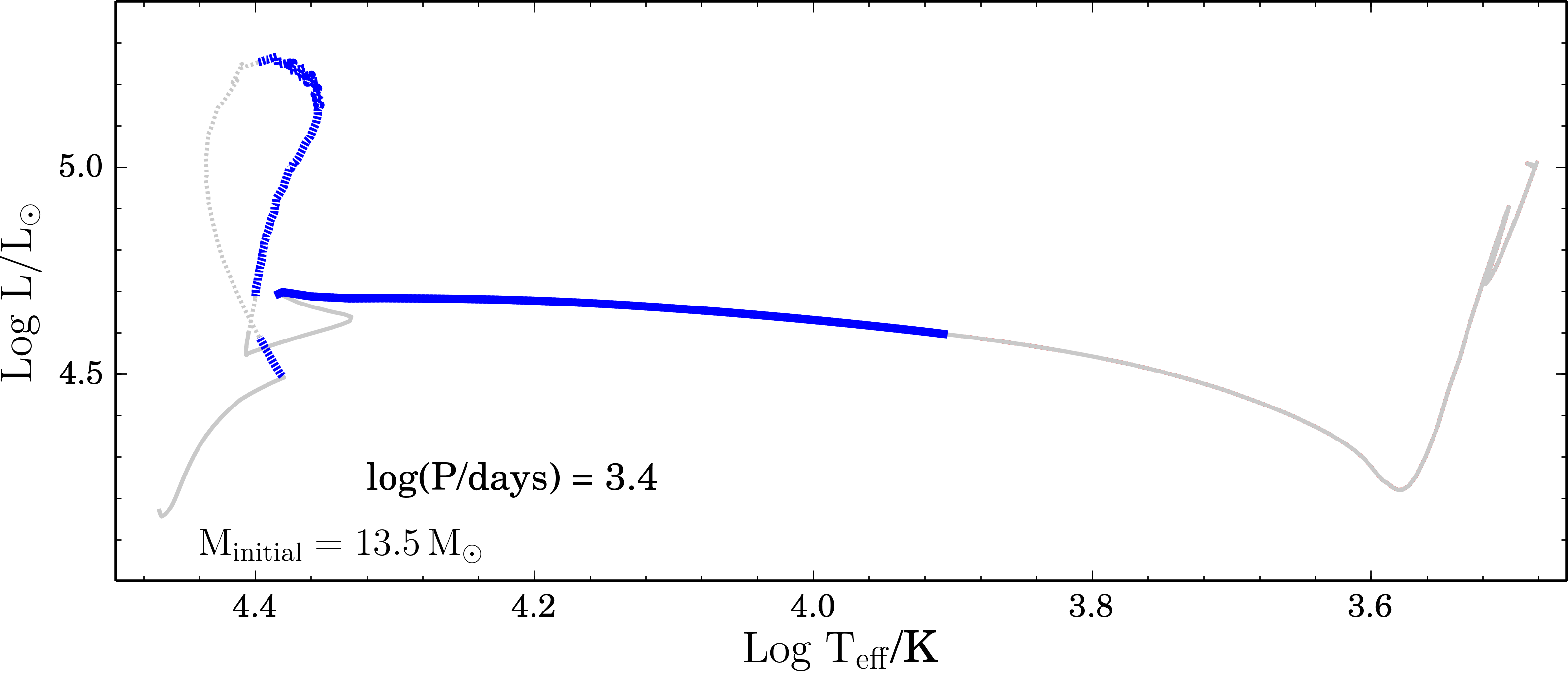}
	\caption{HR diagram for secondary stars in MESA models from the ZAMS to end of carbon burning with initial secondary mass of 13.5 $M_{\odot}$, mass ratio of $q$ = 0.9, Z = 0.020 and an initial orbital period of log($P$/days) = 3.4. The BSG stages are indicated in blue. The dashed part of the track indicates when mass transfer takes place.}
	\label{fig:postSNhrd}
\end{figure}

\begin{figure}
	\centering
	\includegraphics[width=\hsize]{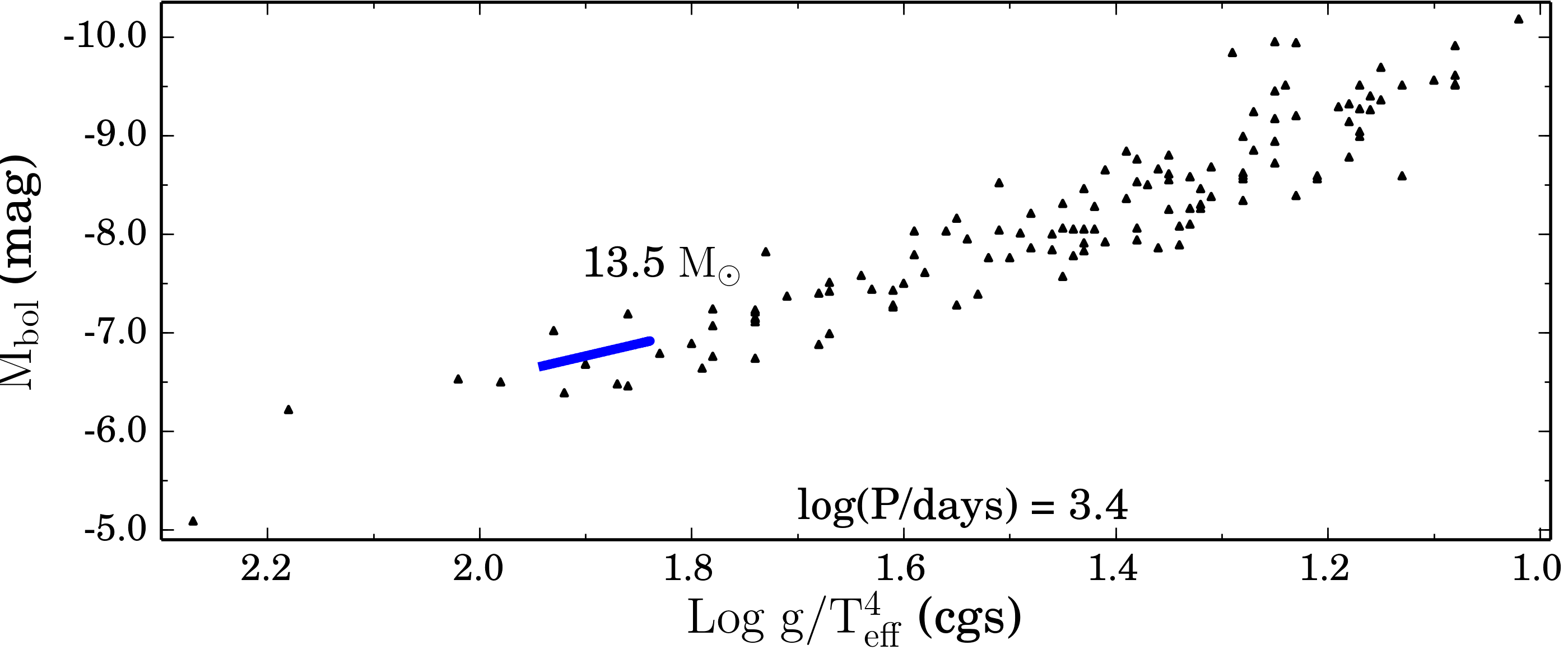}
	\caption{$M_{\rm bol}$ vs. log $g/T_{\rm eff}^4$ plane corresponding to the BSG stage for secondary stars orbiting a compact remnant for the same model as Fig. \ref{fig:postSNhrd}.}
	\label{fig:postSNfwg}
\end{figure}

We compute the remaining evolution of the secondary after the primary explodes for all models with an initial secondary mass of 13.5 $M_{\odot}$. We stop the evolution if the radius of the secondary star is greater than the distance between the two stars (i.e. the secondary star goes into contact with the remnant). This occurs for models with initial periods of log($P$/days) = 1.0 and 2.0.
%For the models with an initial period of log($P$/days) = 1.0 and 2.0, the secondary star and the remnant go into contact.
For the models with larger initial periods, log($P$/days) > 2.0, the separation between the secondary and the neutron star is so large that mass transfer does not occur. This means that the evolution of the secondary, after the explosion of the primary, is somewhat similar to a single 13.5 $M_{\odot}$ star, but with an increased mass and a different internal structure due to previously accreted mass from the primary.
The evolutionary track of a 13.5 $M_{\odot}$ secondary star from zero-age main sequence (ZAMS) to the end of carbon burning is shown in Fig. \ref{fig:postSNhrd} and the track of the secondary (after the explosion of the primary) in the FGLR plane is shown in Fig. \ref{fig:postSNfwg}. 
The location of the tracks in the FGLR plane for the the secondary, after the primary explodes, are in good agreement with the observed FGLR. 
We conclude, at least for these cases under our assumptions, that a secondary that has accreted mass is expected to follow the FGLR.

\begin{table*}[!htbp]
\caption{Selected quantities for primary stars from MESA models taken at log($T_{\rm{eff}}$) = 4.1. `Stage' column denotes if star is moving to the red in the HR diagram (first stage) or moving to the blue (second stage). `f$_{\rm BSG}$' indicates the time spent between log($T_{\rm{eff}}$) = 3.9 and 4.4 for a given stage as a fraction of the total BSG lifetime of each model. $\tau_{\rm BSG}$ indicates the lifetime of the BSG stage in kyr. Not all of these evolutionary points are considered BSGs (see $X$ surface fractions). The rows in italics correspond to models that return to the blue due to mass loss, at a temperature log($T_{\rm{eff}}$) $>$ 3.9 and therefore technically produce only one BSG stage. We include the parameters of these models when it has log($T_{\rm{eff}}$) = 4.1 in the stage 2 rows.} 
\label{table:1}      
\centering
\resizebox{0.9\textwidth}{!}{
\begin{tabular}{c c c || c c c c c c c c || c c c}     
\hline
\multicolumn{2}{c}{Initial details}& & \multicolumn{7}{c}{Quantities for Primary star} & & \multicolumn{3}{c}{Quantities for Secondary star} \\
\hline
$M_{\rm initial}$ & log($P$) & Stage & $M_{\rm prim}$  & f$_{\rm BSG}$ & $\tau_{\rm BSG}$ & $X_{\rm surf}$ & $C/H_{\rm surf}$ & log $g$      & log $P$ & $v^{\rm prim}_{\rm orb}$ & $M_{\rm sec}$  & $L_{\rm pri}/L_{\rm sec}$ & log $g$  \\
$M_{\odot}$   & days   &       & $M_{\odot}$ &      & kyr      &            &               & cm s$^{-2}$ &  days & km s$^{-1}$       & $M_{\odot}$ &  &  cm s$^{-2}$ \\
\hline
12 & 1.0 & 1 & 10.43 & 0.24 & 18.57 & 0.70 & \num{4.92e-03} & 2.70 & 1.06 & 133 & 10.81 & 0.63 & 3.57 \\
12 & 1.0 & 2 & 3.90 & 0.76 & 58.86 & 0.28 & \num{4.52e-04} & 1.68 & 1.84 & 98 & 12.10 & 2.64 & 3.89 \\
\hline 
12 & 2.0 & 1 & 11.59 & 0.31 & 21.96 & 0.70 & \num{4.92e-03} & 2.35 & 2.02 & 60 & 10.57 & 1.98 & 3.63 \\
12 & 2.0 & 2 & 3.98 & 0.69 & 49.91 & 0.31 & \num{3.97e-04} & 1.72 & 2.81 & 46 & 12.08 & 2.74 & 3.86 \\
\hline 
12 & 2.6 & 1 & 11.59 & 0.28 & 22.09 & 0.70 & \num{4.92e-03} & 2.35 & 2.62 & 38 & 10.57 & 1.98 & 3.63 \\
12 & 2.6 & 2 & 4.18 & 0.72 & 56.85 & 0.40 & \num{3.61e-04} & 1.73 & 3.34 & 30 & 12.01 & 2.59 & 3.84 \\
\hline 
12 & 3.0 & 1 & 11.59 & 0.07 & 22.09 & 0.70 & \num{4.92e-03} & 2.35 & 3.02 & 28 & 10.57 & 1.98 & 3.63 \\
12 & 3.0 & 2 & 4.56 & 0.80 & 236.91 & 0.57 & \num{2.29e-03} & 1.95 & 3.60 & 24 & 11.84 & 1.75 & 3.72 \\
\hline 
12 & 3.1 & 1 & 11.59 & 0.04 & 22.10 & 0.70 & \num{4.92e-03} & 2.35 & 3.12 & 25 & 10.57 & 1.98 & 3.63 \\
12 & 3.1 & 2 & 4.72 & 0.96 & 497.86 & 0.61 & \num{2.69e-03} & 2.01 & 3.65 & 23 & 11.76 & 1.58 & 3.66 \\
\hline 
12 & 3.3 & 1 & 11.59 & 1.00 & 22.09 & 0.70 & \num{4.92e-03} & 2.35 & 3.32 & 22 & 10.57 & 1.98 & 3.63 \\
12 & 3.3 & 2 & \multicolumn{7}{l}{does not produce stage that reaches log($T_{\rm eff}$/K) = 4.1} & & & & \\
\hline 
12 & 3.4 & 1 & 11.59 & 1.00 & 22.09 & 0.70 & \num{4.92e-03} & 2.35 & 3.42 & 20 & 10.57 & 1.98 & 3.63 \\
12 & 3.4 & 2 & \multicolumn{7}{l}{does not produce stage that reaches log($T_{\rm eff}$/K) = 4.1} & & & & \\
\hline 
15 & 1.0 & 1 & 10.53 & 1.00 & 50.73 & 0.70 & \num{4.87e-03} & 2.54 & 1.17 & 144 & 14.04 & 0.38 & 3.45 \\
15 & 1.0 & 2 & \textit{5.49} & \textit{1.00} & \textit{50.73} & \textit{0.50} & \textit{\num{2.54e-04}} & \textit{1.72} & \textit{1.72} & \textit{114} & \textit{15.05} & \textit{1.40} & \textit{3.72} \\
\hline 
15 & 2.0 & 1 & 14.42 & 0.31 & 15.81 & 0.70 & \num{4.92e-03} & 2.17 & 2.02 & 65 & 13.30 & 1.83 & 3.54 \\
15 & 2.0 & 2 & 5.76 & 0.69 & 35.78 & 0.32 & \num{3.79e-04} & 1.45 & 2.67 & 54 & 14.99 & 2.66 & 3.75 \\
\hline 
15 & 2.6 & 1 & 14.42 & 0.38 & 15.78 & 0.70 & \num{4.92e-03} & 2.17 & 2.62 & 41 & 13.30 & 1.84 & 3.54 \\
15 & 2.6 & 2 & 5.85 & 0.62 & 25.85 & 0.44 & \num{2.87e-04} & 1.62 & 3.23 & 35 & 14.96 & 2.58 & 3.77 \\
\hline 
15 & 3.0 & 1 & 14.42 & 0.28 & 15.79 & 0.70 & \num{4.92e-03} & 2.17 & 3.02 & 30 & 13.30 & 1.83 & 3.54 \\
15 & 3.0 & 2 & 6.02 & 0.72 & 40.58 & 0.59 & \num{2.13e-03} & 1.73 & 3.55 & 27 & 14.83 & 2.01 & 3.70 \\
\hline 
15 & 3.1 & 1 & 14.42 & 0.18 & 15.79 & 0.70 & \num{4.92e-03} & 2.17 & 3.12 & 28 & 13.30 & 1.83 & 3.54 \\
15 & 3.1 & 2 & 6.13 & 0.78 & 67.92 & 0.61 & \num{2.55e-03} & 1.74 & 3.62 & 25 & 14.76 & 1.84 & 3.68 \\
\hline 
15 & 3.3 & 1 & 14.42 & 0.14 & 15.78 & 0.70 & \num{4.92e-03} & 2.17 & 3.32 & 24 & 13.30 & 1.84 & 3.54 \\
15 & 3.3 & 2 & \multicolumn{7}{l}{does not produce stage that reaches log($T_{\rm eff}$/K) = 4.1} & & & & \\
\hline 
15 & 3.4 & 1 & 14.42 & 1.00 & 15.79 & 0.70 & \num{4.92e-03} & 2.17 & 3.42 & 22 & 13.30 & 1.83 & 3.54 \\
15 & 3.4 & 2 & \multicolumn{7}{l}{does not produce stage that reaches log($T_{\rm eff}$/K) = 4.1} & & & & \\
\hline 
20 & 1.0 & 1 & \multicolumn{7}{l}{does not produce stage that reaches log($T_{\rm eff}$/K) = 4.1} & & & & \\
20 & 1.0 & 2 & \multicolumn{7}{l}{does not produce stage that reaches log($T_{\rm eff}$/K) = 4.1} & & & & \\
\hline 
20 & 2.0 & 1 & 18.73 & 0.39 & 12.41 & 0.70 & \num{4.92e-03} & 1.94 & 2.04 & 71 & 17.69 & 1.80 & 3.42 \\
20 & 2.0 & 2 & 8.02 & 0.30 & 9.39 & 0.48 & \num{4.34e-04} & 1.58 & 2.63 & 61 & 19.83 & 1.47 & 3.84 \\
\hline 
20 & 2.6 & 1 & 18.73 & 0.37 & 11.81 & 0.70 & \num{4.92e-03} & 1.94 & 2.64 & 45 & 17.69 & 1.80 & 3.42 \\
20 & 2.6 & 2 & 8.41 & 0.52 & 16.72 & 0.46 & \num{3.71e-04} & 1.59 & 3.18 & 39 & 19.73 & 1.61 & 3.70 \\
\hline 
20 & 3.0 & 1 & 18.73 & 0.11 & 12.01 & 0.70 & \num{4.92e-03} & 1.95 & 3.04 & 33 & 17.69 & 1.70 & 3.42 \\
20 & 3.0 & 2 & 9.14 & 0.89 & 96.61 & 0.47 & \num{8.58e-04} & 1.59 & 3.46 & 31 & 19.44 & 1.92 & 3.74 \\
\hline 
20 & 3.1 & 1 & 18.73 & 0.13 & 12.01 & 0.70 & \num{4.92e-03} & 1.95 & 3.14 & 30 & 17.69 & 1.70 & 3.42 \\
20 & 3.1 & 2 & 9.07 & 0.87 & 81.64 & 0.51 & \num{1.38e-03} & 1.59 & 3.54 & 29 & 19.35 & 1.76 & 3.65 \\
\hline 
20 & 3.3 & 1 & 18.73 & 1.00 & 12.25 & 0.70 & \num{4.92e-03} & 1.93 & 3.34 & 26 & 17.69 & 1.76 & 3.42 \\
20 & 3.3 & 2 & \multicolumn{7}{l}{does not produce stage that reaches log($T_{\rm eff}$/K) = 4.1} & & & & \\
\hline 
20 & 3.4 & 1 & 18.73 & 0.18 & 10.92 & 0.70 & \num{4.92e-03} & 1.96 & 3.44 & 24 & 17.69 & 1.71 & 3.42 \\
20 & 3.4 & 2 & 8.94 & 0.82 & 51.01 & 0.62 & \num{3.02e-03} & 1.57 & 3.75 & 24 & 18.85 & 1.78 & 3.41 \\
\hline 
30 & 1.0 & 1 & \multicolumn{7}{l}{does not produce stage that reaches log($T_{\rm eff}$/K) = 4.1} & & & & \\
30 & 1.0 & 2 & \multicolumn{7}{l}{does not produce stage that reaches log($T_{\rm eff}$/K) = 4.1} & & & & \\
\hline 
30 & 2.0 & 1 & 25.86 & 1.00 & 48.56 & 0.70 & \num{4.92e-03} & 1.70 & 2.09 & 79 & 25.55 & 1.53 & 3.19 \\
30 & 2.0 & 2 & \textit{15.03} & \textit{1.00} & \textit{48.56} & \textit{0.44} & \textit{\num{4.19e-04}} & \textit{1.39} & \textit{2.42} & \textit{75} & \textit{27.66} & \textit{1.76} & \textit{3.53} \\
\hline 
30 & 2.6 & 1 & 25.86 & 0.09 & 8.07 & 0.70 & \num{4.92e-03} & 1.70 & 2.69 & 49 & 25.55 & 1.56 & 3.19 \\
30 & 2.6 & 2 & 15.83 & 0.91 & 76.96 & 0.49 & \num{3.14e-04} & 1.46 & 2.96 & 48 & 27.35 & 1.73 & 3.43 \\
\hline 
30 & 3.0 & 1 & 25.86 & 0.09 & 8.08 & 0.70 & \num{4.92e-03} & 1.70 & 3.09 & 36 & 25.55 & 1.56 & 3.19 \\
30 & 3.0 & 2 & 16.15 & 0.91 & 80.47 & 0.53 & \num{3.40e-04} & 1.46 & 3.32 & 36 & 27.04 & 1.76 & 3.34 \\
\hline 
30 & 3.1 & 1 & 25.86 & 0.13 & 7.90 & 0.70 & \num{4.92e-03} & 1.68 & 3.19 & 34 & 25.55 & 1.51 & 3.19 \\
30 & 3.1 & 2 & 15.02 & 0.87 & 54.56 & 0.50 & \num{8.64e-04} & 1.40 & 3.45 & 33 & 27.03 & 1.79 & 3.31 \\
\hline 
30 & 3.3 & 1 & 25.86 & 0.14 & 7.93 & 0.70 & \num{4.92e-03} & 1.70 & 3.39 & 29 & 25.55 & 1.54 & 3.19 \\
30 & 3.3 & 2 & 15.46 & 0.86 & 47.79 & 0.46 & \num{9.80e-04} & 1.43 & 3.60 & 29 & 26.55 & 1.78 & 3.22 \\
\hline 
30 & 3.4 & 1 & 25.86 & 0.22 & 8.03 & 0.70 & \num{4.92e-03} & 1.70 & 3.49 & 27 & 25.55 & 1.55 & 3.19 \\
30 & 3.4 & 2 & 15.65 & 0.78 & 29.06 & 0.60 & \num{2.48e-03} & 1.41 & 3.68 & 27 & 26.15 & 1.82 & 3.14 \\
\hline
\end{tabular}}
\vspace{4pt}
\end{table*}

% 30 & 1.0 & 1 & \multicolumn{6}{l}{does not produce stage that reaches log($T_{\rm eff}$/K) = 4.1} & & & & \\

\section{Discussion} \label{sec:discussion}

\subsection{Effect of mass ratio} \label{sec:massratio}
In our analysis above, we assumed a mass ratio of $q$ = 0.9 for the models. We use the outputs from the BPASS models to investigate the effect of the mass ratio, q, on the evolution of the BSGs and the tracks produced in the FGLR. Varying the mass ratio $q$ from 0.1 to 0.9, for the same primary mass and initial period, we find the tracks in the FGLR plane are almost coincidental, indicating that the mass ratio of the system has very little effect on the predicted FGLR sequence. Some of these systems may go into contact, especially those with a small mass ratio, for a given initial orbital period and an initial mass of the primary. However as we have not modelled such scenarios, we cannot say much further about these systems. These systems are potentially important as previous studies have indicated that binary systems with short initial periods resulting in Case B mergers are a promising channel for BSG production \citep{1992ApJ...391..246P, 2014ApJ...796..121J, 2017MNRAS.469.4649M}.

\subsection{Range of orbital Periods leading to BSGs}
The range of initial orbital periods we study is limited to 100 -- 2,500 days. By studying the BPASS suite of models for larger initial periods, we find that primary stars of mass 30 $M_{\odot}$ in binary systems with initial periods $P$ $\gtrsim$ 4,000 days (log($P$/days) = 3.6) will not undergo mass transfer to the secondary. Therefore, while some binary systems will have  periods $\gtrsim$ 4,000 days, the stars will not exchange mass due to RLOF and will evolve similarly to single stars.
The discussion of the shorter period system is more complicated because different physical processes become important. At short distances, tidal interactions occur and systems are more prone to go into contact during their evolution. 
Although these effects are very important for the evolution of these stars, they will likely not produce bona fide BSGs, because they may induce chemically homogeneous evolution (CHE) and thus move the stars away from the BSG region of the HR diagram \citep{2016A&A...585A.120S, 2010ApJ...725..940Y, 2009A&A...497..243D}. To induce CHE by tidal interaction, a very short initial orbital period is needed and therefore the parameter space for CHE is rather small \citep{2016A&A...585A.120S}. Also, at solar metallicity, strong stellar winds tend to widen the orbit, which can prevent CHE. While a large fraction of Case A mass transfer systems would also produce BSGs \citep[see, for example][]{2010ApJ...725..940Y}, the parameter space for Case A systems is much smaller than for Case B systems and therefore we expect that Case A systems would only have a minor contribution to the production of BSGs.
The Case A fraction depends on how large stars become on the main sequence and this is influenced by both rotation and uncertain amounts of extra-mixing in stellar models.

\subsection{Hydrogen abundance at the surface}
The minimum hydrogen surface fraction X, that we assume in our definition of a BSG stage, significantly affects the fraction of stars that produce a second BSG stage. For this work, we define the BSG stage with $X$ > 0.5. Decreasing or increasing the minimum $X$ surface abundance results in a respective increase or decrease in the fraction of stars that produce a second BSG stage. For example, following the same method as in Sect. \ref{sec:secondstagebsg}, assuming a BSG for $X$ > 0.3 (X > 0.6) predicts 24\% (2\%) of 15 -- 30 $M_{\odot}$ primary stars to be located away from the tight scatter in the FGLR plane. The evolution as a function of time of the hydrogen surface fractions depend on the stellar models, and in particular on how mass loss and mixing are treated.

In principle, the fraction of outliers could be used to constrain the models if a complete observational sample were available. However we note that the target selection for spectroscopic FGLR distance determinations is heavily biased towards brighter objects to enable spectroscopy with a decent signal-to-noise. To assess the statistical effect of this selection bias on the fraction of outliers is difficult without population synthesis. 
The observations suggest a very small fraction of such objects. For nearby galaxies with distances smaller than 2 Mpc (M33, WLM, NGC3109, NGC300), only two objects out of a total of 81 (or 2.5\%) where found to be low mass FGLR outliers.

%in principle the fraction of outliers oculd be used to constrain the models if a complete observational sapmle were available. However we note that the target selection is baised towards brigher objects to enable..... noise. then pop syn setence
%also quote 2.5% figure
% \textbf{The observations suggest a very small fraction of such objects. For nearby galaxies with distances smaller than 2 Mpc (M33, WLM, NGC3109, NGC300), only two objects out of a total of 81 (or 2.5\%) where found to be low mass FGLR outliers. However, we note that the target selection for spectroscopic FGLR distance determinations is heavily biased towards brighter objects to enable spectroscopy with a decent signal-to-noise. To assess the statistical effect of this selection bias on the fraction of outliers is difficult without population synthesis.}

%\subsection{Difference between BPASS and MESA}
%Maybe I should just put this in earlier...

\subsection{Impact of stellar rotation}
In this subsection, we discuss how stellar rotation and its effects on mixing and core size may affect the formation and properties of BSGs. \citet{2015A&A...581A..36M} compare the observed FGLR to single star models with and without rotation. They conclude that single star models with rotation showed a slightly better agreement with the observed FGLR than those without rotation. We chose to compute all the MESA models in this study without rotation in order to isolate the effect of mass transfer via RLOF on the FGLR. For long period binary systems, we would expect the inclusion of rotation in the models to have the same effect on the FGLR sequences as in the case of single stars and hence a slightly better reproduction of the observed FGLR. In very short period systems (with an orbital period of the order of one day), tides rapidly cause the period of rotation of the star to become equal to the orbital period (a process called synchronisation). Since the orbital period is short, stars may be rotating so quickly that they may follow a homogeneous evolution \citep{2016A&A...585A.120S, 2010ApJ...725..940Y, 2009A&A...497..243D}. Stars in these systems will likely not evolve to a BSG.

Rotation has an impact of the duration of the RSG phase and thus impacts the probability that a system will undergo RLOF during that phase. This will have further consequences on the evolution of the star. Some effects of rotation, such as the increase of the mass of the core during the MS phase, favour a rapid redward evolution after the MS phase causing the beginning of the RSG phase at an early stage of the core He-burning process. This may favour RLOF during the RSG phase. Other processes, such as the mixing of He into the H-rich envelope, has mixed effects that can both favour and disfavour a rapid redward evolution after the MS phase. On one hand, helium mixing into the external H-rich layers reduces the mass fraction of hydrogen in the H-burning shell. This tends to decrease the efficiency of the H-burning shell, to reduce the size of the intermediate convective zone attached to it and to favour a rapid redward evolution \citep{2001A&A...373..555M}. On the other hand, the helium mixing makes the star more homogeneous, tending to keep it in a bluer region of the HR diagram. This disfavours the evolution into a RSG phase, at least at an early stage of the core He-burning phase \citep{2013LNP...865.....G}.

It is worth noting that the treatment of convection also plays a crucial role in producing BSGs, for instance using the Ledoux criterion instead of the Schwarschild criterion makes a difference as discussed by \citet{2014MNRAS.439L...6G}. The choice of the convection criterion also affects the surface chemical composition and possibly the timing of the mass transfer episodes in binary systems.

This interplay between the mass of the core, the behaviour of the convective zone associated to the H-burning shell and the degree of overall mixing is complex and remains to be more thoroughly explored. Rotation does affect the duration of the RSG phase and thus, at least in an indirect way, any RLOF that will occur at that stage. 
At the moment, the quantitative effects of rotation on the properties of BSGs that appear after the RSG phase is still an open question.
However, it is difficult to say more quantitatively how it affects the properties and the frequency of the stage 2 BSGs. This question remains largely open.

% -- Photometry and spectroscopy --
\subsection{Impact of an unresolved secondary on photometry and spectroscopy of a BSG primary star} \label{sec:unresol_dis}
For the range of initial orbital periods chosen in the binary models produced using MESA, the typical separation between the primary and secondary stars during the BSG stage ranges from  200 -- 1200 $R_{\odot}$. 
%Using a 10 m telescope, it is not possible to resolve the two stars beyond 400 pc at visible wavelengths. 
As the FGLR is used as an extragalactic distance indicator with BSGs at distances of the order of \textasciitilde Mpc, the primary and secondary stars are unresolved at these distances.

Because of this, it is important to study the impact of the presence of an unresolved secondary star on the observed quantities obtained for the primary star. It is possible that the presence of a secondary companion will contaminate the spectrum and affect quantities derived from the spectrum such as the $B-V$ colour used for reddening corrections, the value obtained for log $g$ or for $T_{\rm eff}$. The increased flux from an unresolved secondary may also contribute to the bolometric magnitude assigned to the primary.

In the context of post-interaction binary systems, \citet{2017A&A...608A..11G, 2018arXiv180203018G} looked at the detectability of stripped stars and found that they may be challenging to detect at optical wavelengths, but easier to detect at UV wavelengths. Here, we look at the pre-interaction detectability of binary systems and in particular how the flux from a secondary may impact the spectrum of a primary BSG.

\subsubsection{Determination of log $g$ and $T_{\rm{eff}}$}
% Log g
\begin{figure}
	\centering
	\includegraphics[width=\hsize]{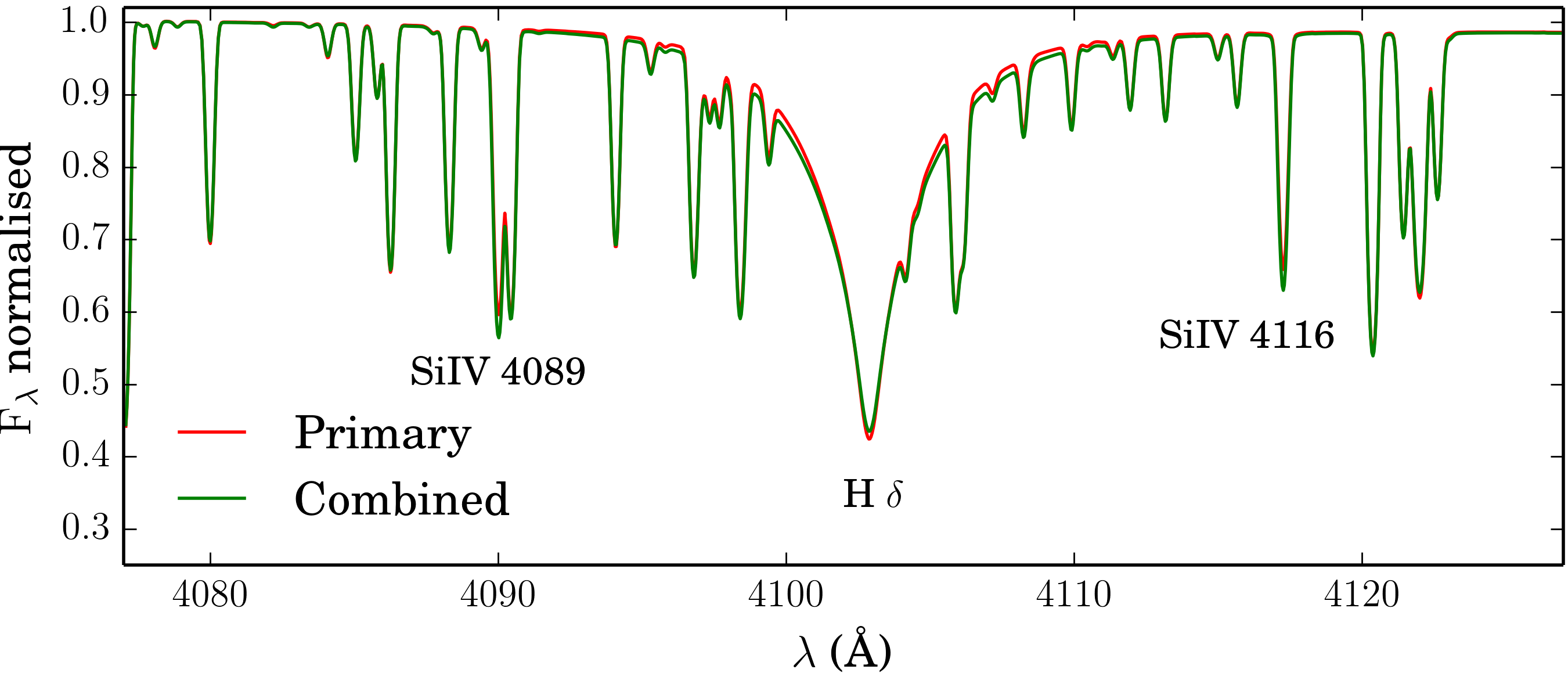} \includegraphics[width=\hsize]{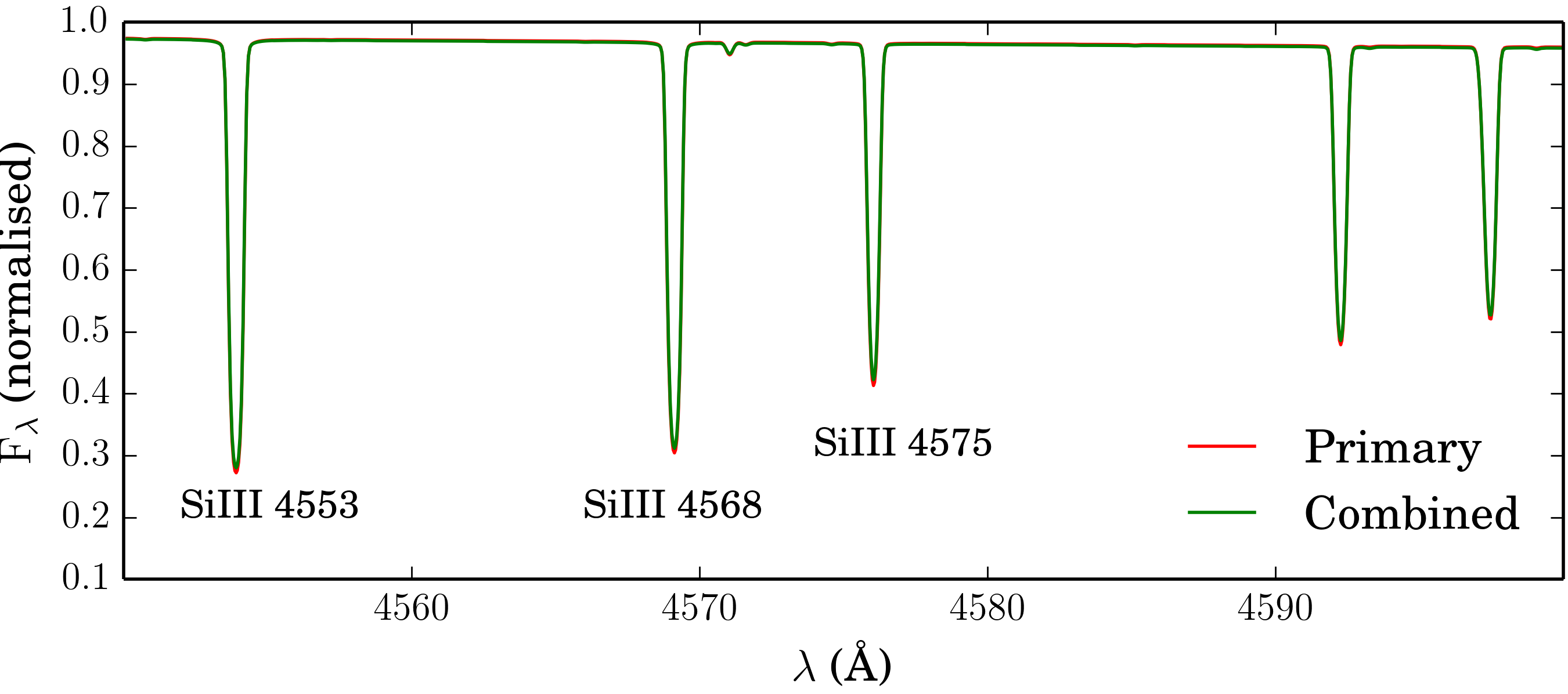}
	\caption{CMFGEN model spectrum computed at the beginning of the BSG stage using outputs from our MESA binary models for a representative system of a 20 $M_{\odot}$ BSG primary and an 18 $M_{\odot}$ (MS) secondary. The red line profiles indicate the CMFGEN model spectrum of a primary star with T = 23\,000~K and log(g) = 3.0 dex. The green line indicates a combination of the model spectrum of the primary star with T = 23\,000~K and log(g) = 3.00 dex and the secondary star with T = 26\,000~K and log(g) = 3.50 dex. The flux ratio of the primary to the secondary in the B band at the stage when the spectra are computed is $F_{B, \, \mathrm{pri}}/F_{B, \, \mathrm{sec}}$ = 2.3. {\it Top panel:} 
Spectral region around the H$\delta$ line which is one of the diagnostics for log $g$.
{\it Bottom panel:} Spectral region around SiII lines at 4553, 4568 and 4575 {\AA}, which are $T_{\rm eff}^4$ diagnostics along with SiII and SiIV lines.}
	\label{fig:hlines}
\end{figure}

\begin{figure}
	\centering
	\includegraphics[width=\hsize]{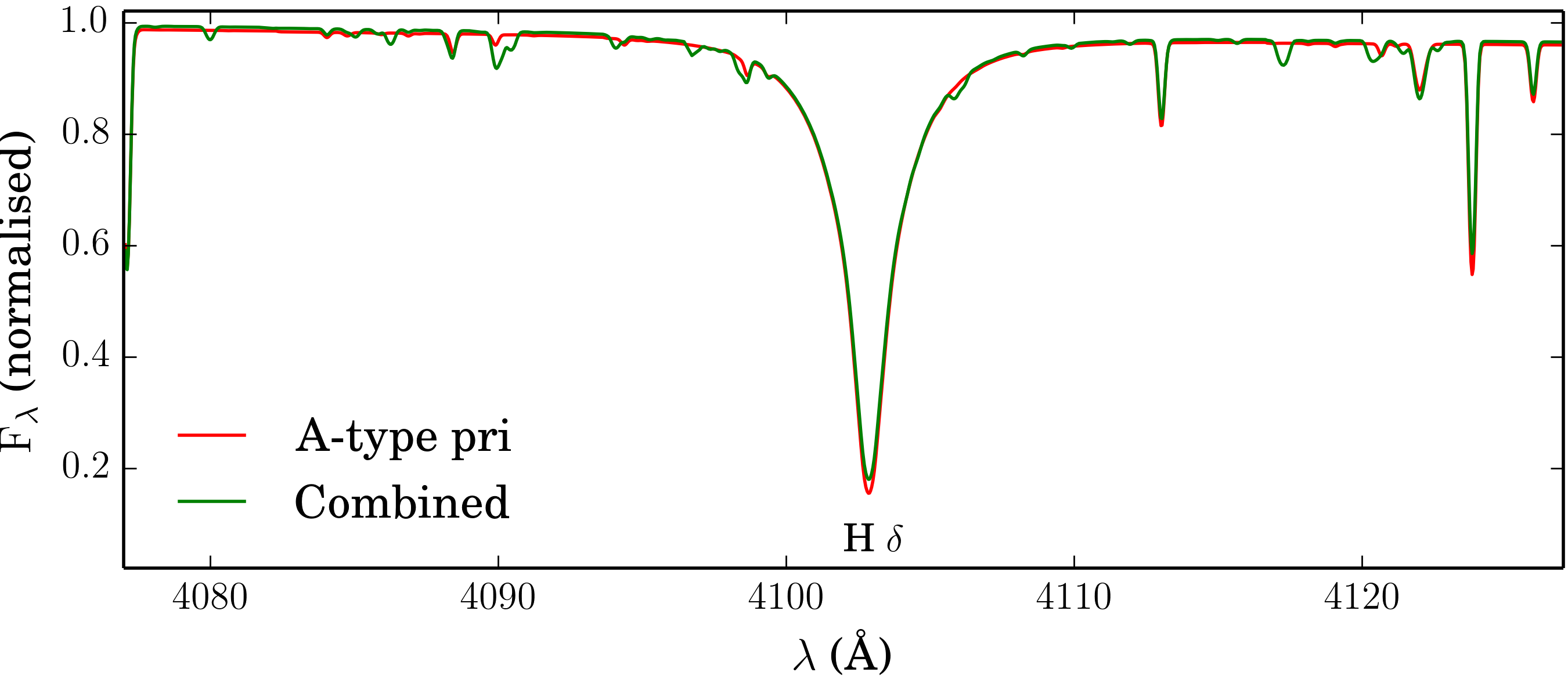} \includegraphics[width=\hsize]{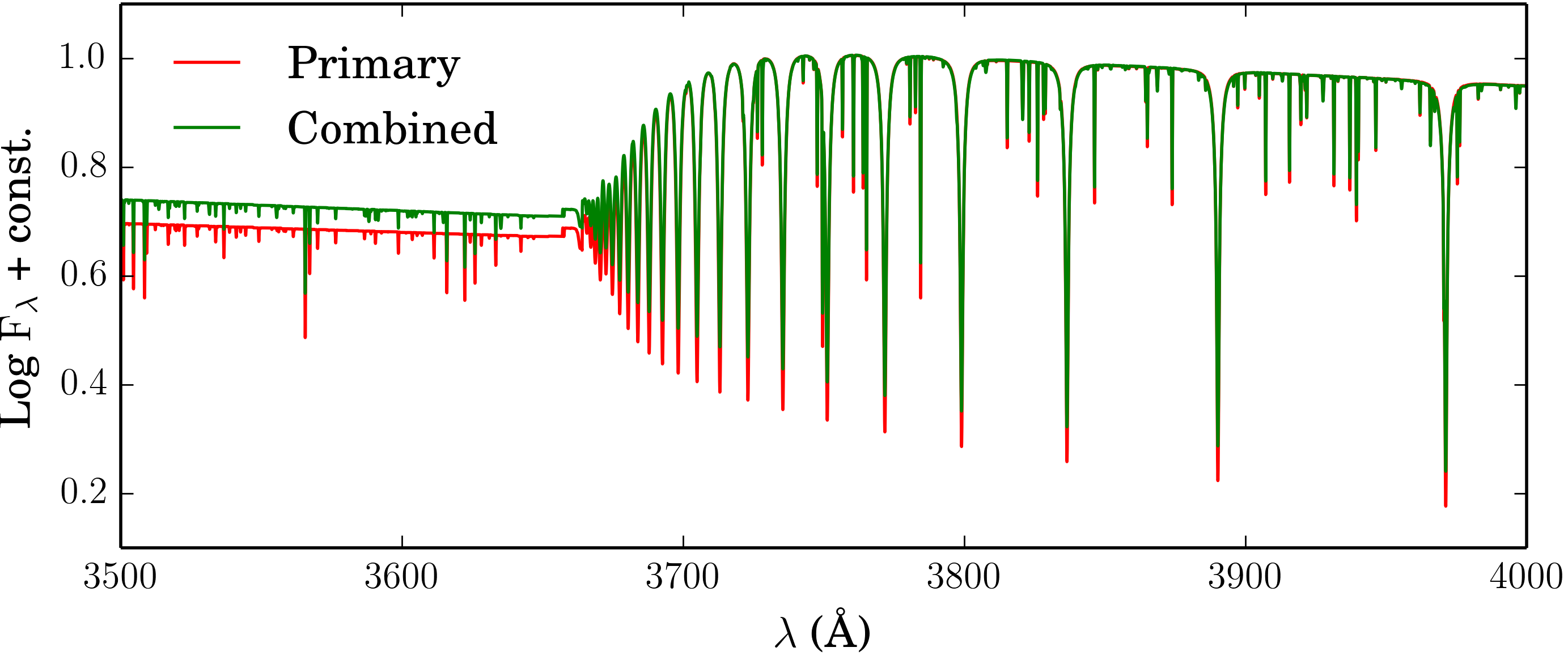} \includegraphics[width=\hsize]{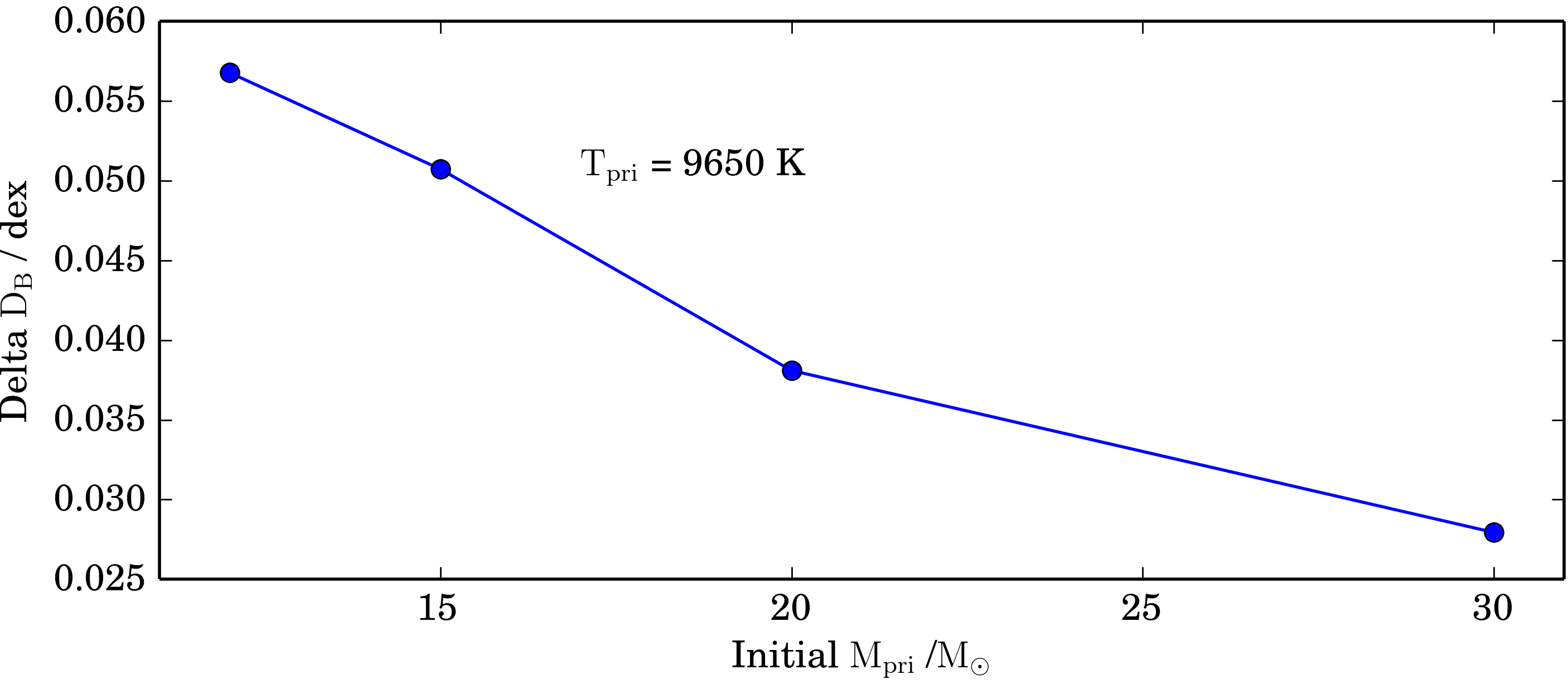}
	\caption{CMFGEN model spectra computed at the end of the BSG stage using outputs from our MESA binary models, taken from the same binary models as in Fig. \ref{fig:hlines}, but at a later evolutionary stage. 
	The red line profiles indicate the CMFGEN model spectrum of a primary star with T = 9\,650~K and log(g) = 1.50 dex. The green line indicates a combination of the model spectrum of the primary star with T = 9\,650~K and log(g) = 1.50 dex and the secondary star with T = 26\,000~K and log(g) = 3.50 dex. The flux ratio of the primary to the secondary below the Balmer jump is 3.0. {\it Top panel:} H$\delta$ line for the primary (red) and combined spectra (green). {\it Middle panel:} Balmer jump for the primary (A-type supergiant; red) and combined spectrum (green). Fluxes are normalised at 3790 {\AA} for clarity. {\it Bottom panel:} Change in Balmer jump in the combined spectrum relative to the primary spectrum, as a function of initial primary mass. The combined spectrum has the lower Balmer jump.}
	\label{fig:hlinesatype}
\end{figure}

We first look at how the presence of an unresolved secondary would affect the values obtained from the spectrum for log $g$ and $T_{\rm eff}$. 

%When calibrating and applying the FGLR, the higher Balmer lines are used to determine values for log $g$ and ionisation equilibria (Si II/III/IV, supplemented by others e.g. He I/II) are used to determine temperatures \citep[e.g.][]{2017AJ....154..102U}. Temperatures and values for log $g$ of A-type BSGs can also be obtained from the Balmer jump.
In the analysis of observed BSG spectra, stellar gravities are constrained through a model atmosphere fit of the higher Balmer lines. For the determination of effective temperatures, different methods are applied depending on the spectral type. For BSGs of spectral type B0 to B5, the ionisation equilibrium of Si II/III/IV is used \citep[e.g.][]{2005ApJ...622..862U}. For later spectral types (B6 to A4), $T_{\rm eff}$ was originally obtained from a fit of the Balmer jump in the earlier FGLR work \citep[see][for details]{2008ApJ...681..269K}, however this method has since been replaced by a $\mathrm{\chi^2}$ fit of the total metal line spectrum between 4000 and 5500 {\AA} which constrains $T_{\rm eff}$ and metallicity simultaneously \citep{2014ApJ...785..151H, 2013ApJ...779L..20K}. To study the effect of the secondary on these spectral features, we select a representative case from our MESA binary models ($M_{\rm pri}$ = 20 $M_{\odot}$, $M_{\rm sec}$ = 18 $M_{\odot}$, log($P$/days) = 3.0) and use the radiative transfer code CMFGEN \citep{1998ApJ...496..407H} to compute synthetic spectra of a primary and secondary at the beginning (Fig. \ref{fig:hlines}) and the end (Fig. \ref{fig:hlinesatype}) of the first BSG stage of the primary (before mass transfer takes place).

We compute spectra when the primary has a temperature of $T_{\rm eff}$ = 23,000~K with log(g) = 3.00 dex (beginning of BSG stage) and when the primary has a temperature of $T_{\rm eff}$ = 9,650~K with log(g) = 1.50 dex (end of BSG stage). The secondary has a temperature of $T_{\rm eff}$ = 26,000~K with log(g) = 3.50 dex in both cases. 
The radiative transfer models are similar to those described in \citet{2014A&A...564A..30G} and \citet{2017MNRAS.468.2333S}.
The CMFGEN spectra we compute in this paper are based on BSG models presented in \citet{2017MNRAS.468.2333S} and YSG models presented \citet{2014A&A...564A..30G}, using a similar atomic model. In this paper, we computed a small grid of models around the values of log $g$, $T_{\rm eff}$ and luminosity predicted by the binary models for the primary and secondary with  $\dot{M}$ similar to that of the MESA models. The abundances are also the same as those from the MESA models at the appropriate evolutionary state.
%($\dot{M}$ is low so it does not greatly affect the optical diagnostic lines or the Balmer jump)
We note that this choice of mass ratio ($q$ = 0.9) and $T_{\rm eff}$ of the primary represent a `worst case scenario' (i.e. maximising the relative contribution of flux from the secondary in the V band). We would expect that other binary systems containing a 20 $M_{\odot}$ BSG would show a lower contamination due to the secondary. We combine the spectrum of the primary and secondary using their luminosities and compare the combined spectrum to that of the primary to study the effect of the presence of the secondary. 

As an example of a Balmer line used to determine log g, we compare the H$\delta$ line for the primary and combined spectra (top panels of Figs. \ref{fig:hlines} and \ref{fig:hlinesatype}). At the beginning of the BSG stage, a small amount of increased broadening in the H$\delta$ line is noticeable in the combined spectrum due to the higher log $g$ of the secondary. 
However, this difference is too small to significantly affect the log $g$ determination, especially at the spectral resolution of 5 {\AA} which is used for the extragalactic studies of BSGs. The effect on the Si III/IV lines is also very small which suggests that the presence of a secondary has little influence on the temperature and gravity diagnostics of early BSG types. At the end of the BSG stage, when the primary has $T_{\rm eff}$ = 9\,650~K, the Balmer lines and the metal lines, and hence the determination of log $g$ and $T_{\rm eff}$, are practically unaffected by the presence of the secondary (upper panel of Fig. \ref{fig:hlinesatype}). This is because the ratio of the flux in the  V-band of the primary to the secondary is 16 (see Sect. \ref{sec:discusflux}).

We also investigate the temperature diagnostic for later spectral types using the Balmer jump, as it has been applied in the earlier FGLR work. In Fig. \ref{fig:hlinesatype}, we compare the Balmer jump for the primary star with the combined spectral energy distribution at the end of the BSG stage when the primary is an A-type BSG. At wavelengths lower than the Balmer jump, the flux ratio of the primary to the secondary is 3.0 because of the higher $T_{\rm eff}$ of the secondary. As a result, the Balmer jump of the combined spectral energy distribution is slightly decreased, as compared the the Balmer jump of the primary. In the bottom panel of Fig. \ref{fig:hlinesatype}, we plot the difference in Balmer jump inferred from the primary and combined spectra, $\Delta$ DB, for models with different initial primary masses by computing CMFGEN spectra for the primary when $T_{\rm pri}$ = 9\,650~K and for secondaries of different masses. The value of $\Delta$ DB increases with increasing initial mass because the values for log $g$ of the primary at $T_{\rm pri}$ = 9\,650~K decrease with increasing mass. The lower log $g$ of the primary then results in a smaller value of DB of the primary which means the UV flux of the secondary is less important when the energy distributions of the primary and secondary are combined. As can be inferred from Fig. 30 in \citet{2008ApJ...681..269K}, this reduced Balmer jump will result in an increase of $T_{\rm eff}$ of up to 300~K and also an increase of $g_{\rm F}$ of up to 0.05 dex. This is within the uncertainties of the temperature and gravity diagnostics, but it will be a systematic effect as the secondaries are always hotter than the primary in our models with $q = 0.9$. As noted at the beginning of this section, this is the most extreme case. Population synthesis, taking into account the distribution of mass fractions and initial orbital periods will be needed to assess this effect in detail. 

We conclude that there is a small systematic bias in the determinations of log $g$ and $T_{\rm eff}$ from spectra of primary BSGs in binary systems due to the presence of unresolved secondary stars.

\begin{figure}
	\centering
	\includegraphics[width=\hsize]{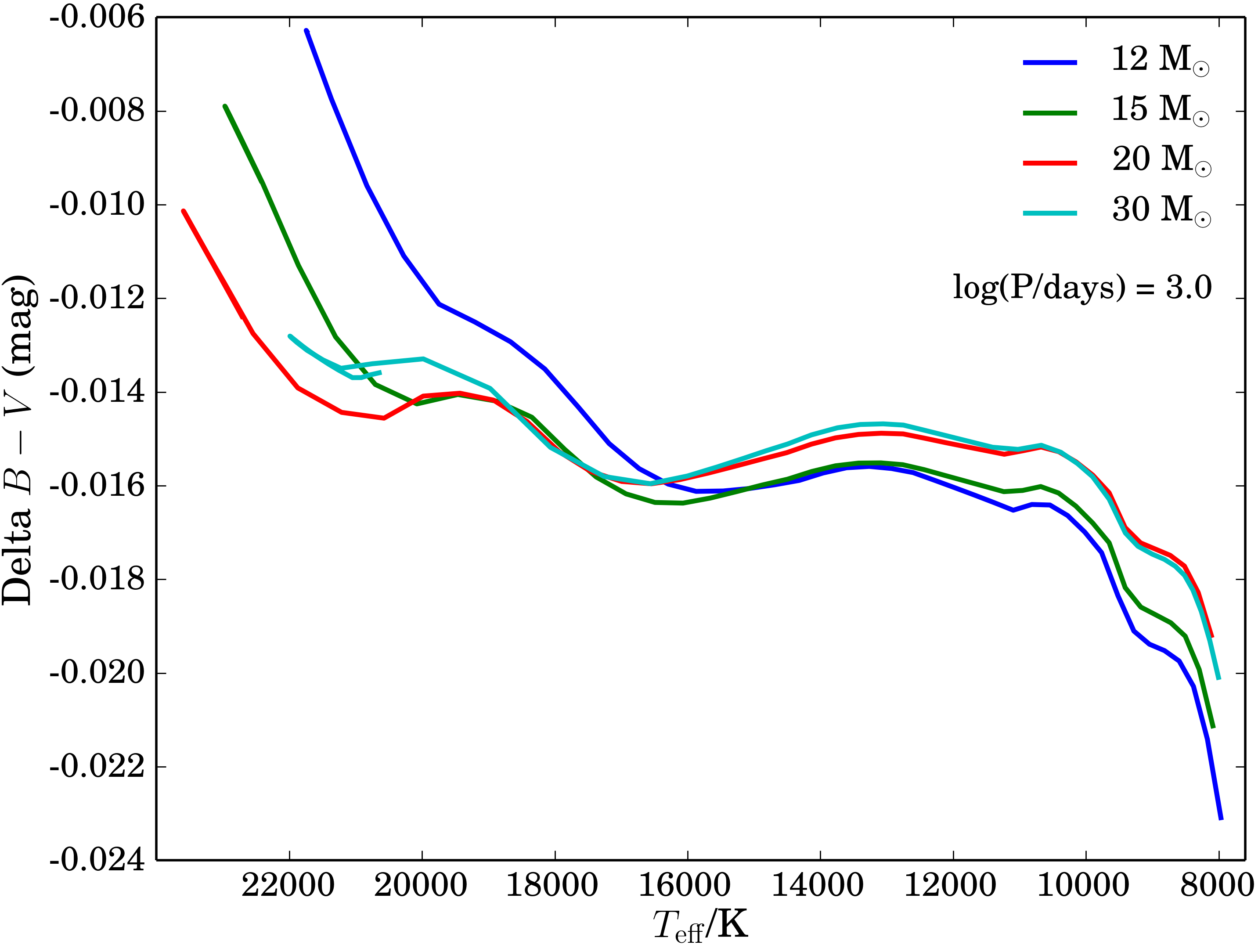}
	\caption{Change in $B-V$ colour due to presence of secondary as a function of the effective temperature of the primary star during the first BSG stage of the primary (before mass transfer takes place). The values were computed for our MESA models with initial primary masses of 12, 15, 20 and 30 $M_{\odot}$, log($P$/days) = 3.0 and mass ratio $q$ = 0.9.}
	\label{fig:deltabv}
\end{figure}

\subsubsection{Determination of $B-V$ colour}
% B-V colour
Secondly, we discuss how the presence of an unresolved secondary may affect the $B-V$ colour used to calculate reddening corrections.
As the secondary has a temperature of about $T_{\rm eff} = 26,000 $K throughout the BSG stage of the primary, its relative contributions in the B and V band may cause a hotter inferred $B-V$ colour. A changed $B-V$ colour due to the secondary may result in incorrect reddening corrections and hence incorrect bolometric magnitudes assigned to the primary.
To check this, we plot the change in $B-V$ that would be observed due to the presence of the secondary star using colours from \cite{2011ApJS..193....1W}, as a function of $T_{\rm eff}$ of the primary star during the first BSG stage (Fig. \ref{fig:deltabv}). The flux contribution of the secondary has little effect on the $B-V$ colour of the unresolved system, causing a maximum change in $B-V$ colour of $-0.023$ mag. This is also encouraging because it shows that the reddening corrections are not strongly affected by the presence of the secondary. While this is a small effect, it is also systematic. A systematic error of 0.02 mag in $B-V$ would result in an error of 0.06 mag in extinction and, thus, in the apparent bolometric magnitude $M_{\rm bol}$. As a result, the distance modulus determined would be slightly biased towards smaller values. This implies that the FGLR method applied to BSGs that are in unresolved binaries would slightly underestimate the distance to their host galaxies. However, we note again that the mass ratio considered here $q$ = 0.9 is the most extreme case.
%and that population synthesis would be needed to investigate and quantify the significance of systematic effect.

\subsubsection{Determination of the bolometric magnitude} \label{sec:discusflux}
%Mbol introduction
It is also important to check how the increased flux from an unresolved secondary affects the bolometric magnitude assigned to the primary and the consequences of this increased flux on the FGLR. As described in \citet{2008ApJ...681..269K}, the effective temperatures, gravities and metallicities obtained in the spectral analysis of each individual BSG are used to calculate bolometric corrections (BCs), which are then combined with the de-reddened observed V-band magnitudes to obtain apparent bolometric magnitudes $M_{\rm bol}$. Therefore, the secondary can affect $M_{\rm bol}$ either due to additional flux in the V band, or due to spectral contamination causing an incorrect determination of log $g$ or $T_{\rm eff}$, which are used to calculate the BCs. As we have discussed above, the determinations for log $g$ and $T_{\rm eff}$ of the primary BSG are not significantly affected by the presence of an unresolved secondary. However, increased flux in the V band from an unresolved secondary may cause an increase in the calculated bolometric luminosity and an increased scatter of the FGLR. This increase in apparent bolometric luminosity may result in underestimates of distances to BSGs.

% --- Fv ratios ---

\begin{figure}
	\centering
	\includegraphics[width=\hsize]{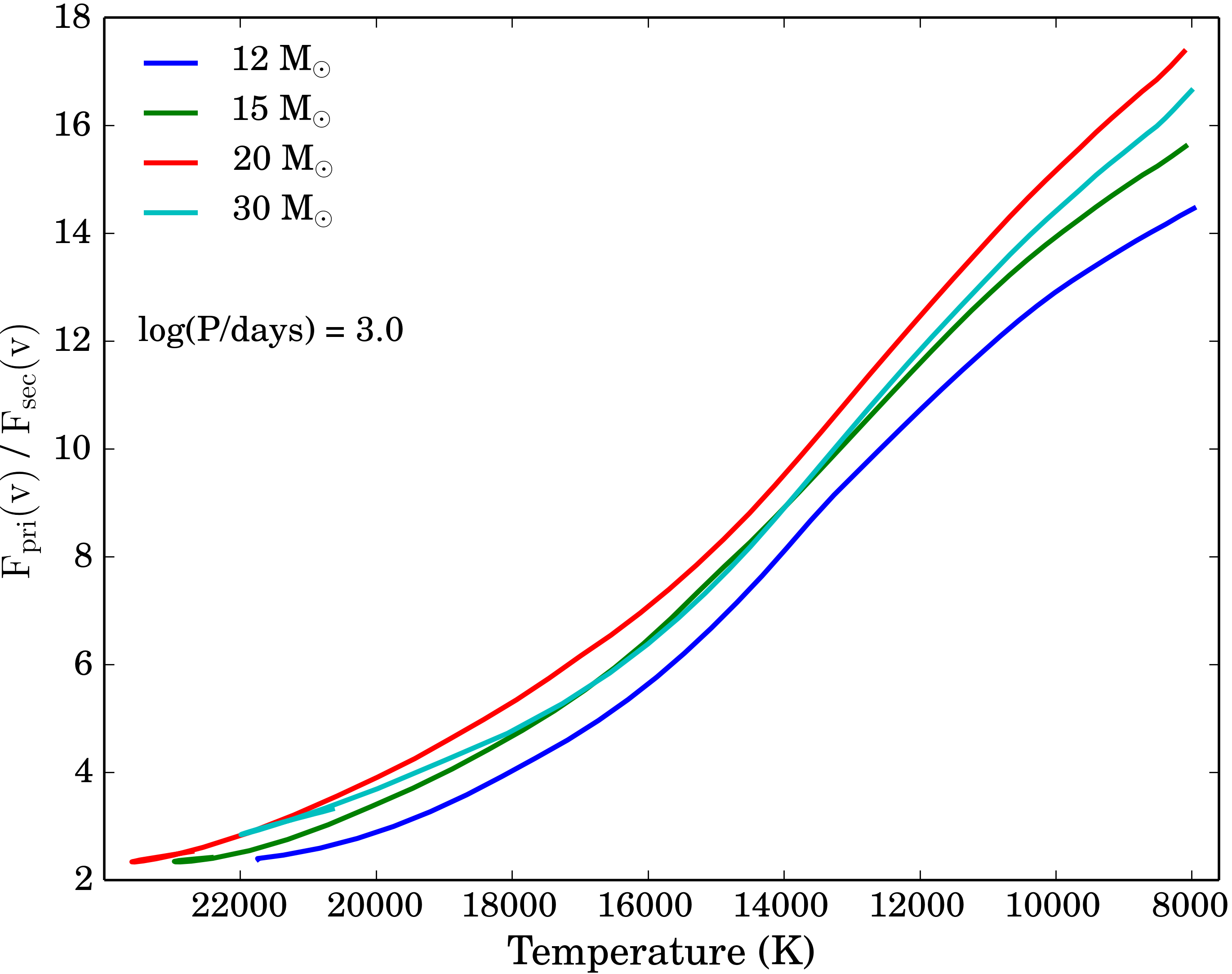}
	\caption{The ratio of the flux of the primary to the secondary in the V band, $F_{V, \, \mathrm{pri}}/F_{V, \, \mathrm{sec}}$, as a function of the temperature of the primary star during the main BSG stage of the primary (before mass transfer takes place). The values were computed for our MESA models with initial primary masses of 12, 15, 20 and 30 $M_{\odot}$, log($P$/days) = 3.0 and mass ratio $q$ = 0.9.}
	\label{fig:fvratios}
\end{figure}

To study the effect of increased flux in the V band, we investigate the ratio of the flux in the V band of the primary to the secondary, $F_{V, \, \mathrm{pri}}/F_{V, \, \mathrm{sec}}$, during the main BSG stage of the primary (Fig. \ref{fig:fvratios}). We select some representative models to illustrate the behaviour for other systems. Again, we note that this mass ratio of $q$ = 0.9 represents a worst case scenario in terms of contribution of flux from the secondary. During the BSG stage, the bolometric luminosities of both the primary and secondary and the temperature of the secondary remain approximately constant over the timescale of the BSG stage. The value of $F_{V, \, \mathrm{pri}}/F_{V, \, \mathrm{sec}}$ increases as the temperature of the primary decreases and its output in the V band increases. For much of the BSG stage of the primary, the flux ratio of the primary to the secondary is so large that the secondary will have a negligible impact on the total flux in the V band or on the spectrum in the visual range. This means that even for a BSG in a binary system with mass ratio $q$ = 0.9, the presence of an unresolved secondary will only impact the V band flux from the system during the beginning of the BSG stage of the primary. For systems with lower mass ratios, we expect much larger flux ratios $F_{V, \, \mathrm{pri}}/F_{V, \, \mathrm{sec}}$.

\begin{figure}
	\centering
	\includegraphics[width=\hsize]{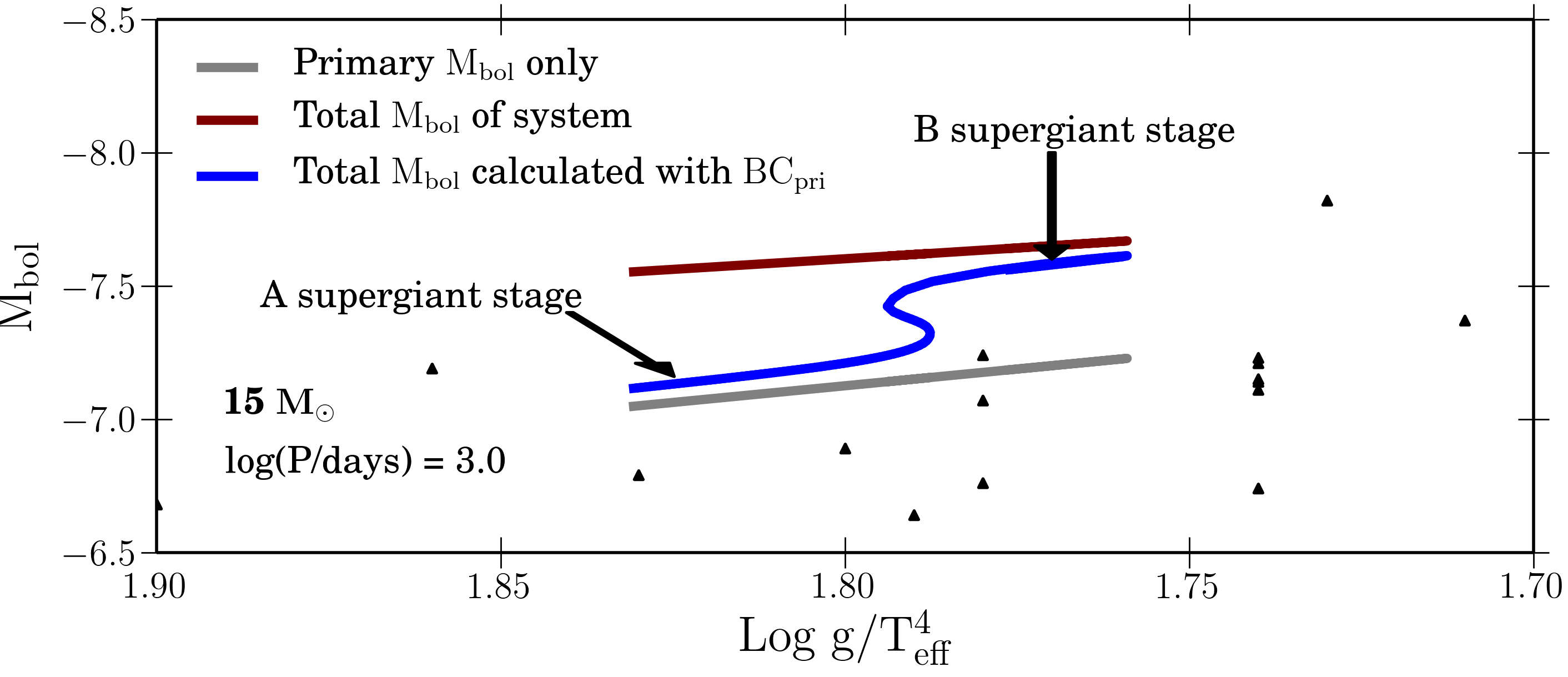} \includegraphics[width=\hsize]{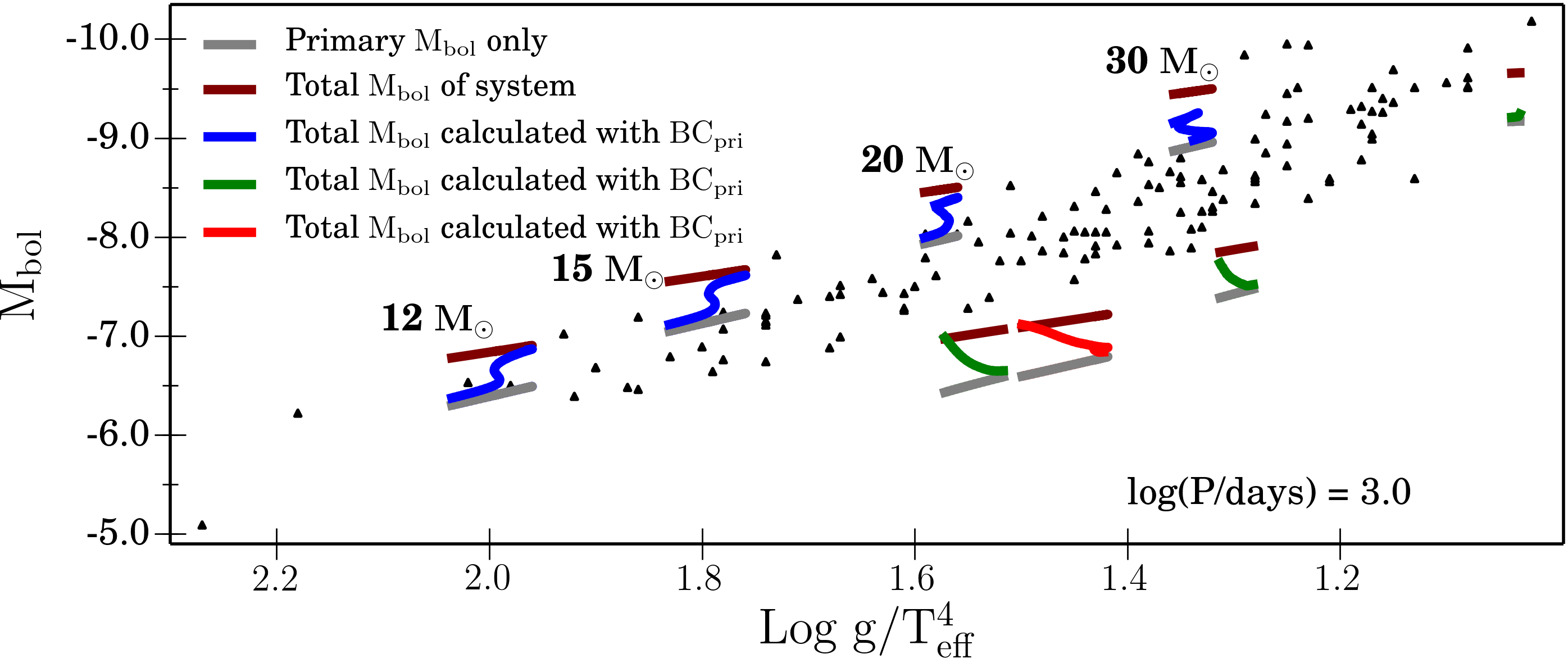}
	\caption{Tracks in $M_{\rm bol}$ vs. Log $\mathrm{g/T^4_{eff}}$ plane for BSG stages, using both a modified and unmodified bolometric magnitude due to the presence of the secondary. Grey tracks are original tracks with unmodified $M_{\rm bol}$ (as in Fig. \ref{fig:mesaprimmodels}). Blue, green and red tracks correspond to the first, second and third BSG stages with modified $M_{\rm bol}$ calculated by combining the V band flux of the primary and secondary and applying the bolometric correction (BC) based on the log $g$ and $T_{\rm eff}$ of the primary. Maroon tracks use a modified $M_{\rm bol}$ calculated by computing the bolometric magnitude for the sum of the luminosities of the primary and secondary. {\it Top panel:} Close-up track of the 15 $M_{\odot}$ model to more clearly illustrate the changes in the modified tracks. {\it Bottom panel:} Models with initial masses of 12, 15, 20 and 30 $M_{\odot}$, mass ratio of $q$ = 0.9 and initial period of log($P$/days) = 3.0.}
	\label{fig:fwgmodified}
\end{figure}

%Modified FGLR planes
Based on the conclusions above that there is little systematic bias in determinations of log $g$ and $T_{\rm eff}$ from spectra of primary BSGs in binary systems, we expect that the the presence of unresolved secondary stars will not shift the tracks in the FGLR diagram horizontally. However, the analysis of the values of $F_{V, \, \mathrm{pri}}/F_{V, \, \mathrm{sec}}$, suggests that changes in $M_{\rm bol}$ assigned to the primary may shift the tracks vertically in the FGLR diagram.
To study the effect of the presence of a secondary star on the $M_{\rm bol}$ assigned to the primary on the FGLR, we compute tracks in the $M_{\rm bol}$ vs. Log $\mathrm{g/T^4_{eff}}$ plane for a representative sample of MESA models in Fig. \ref{fig:fvratios}, using a modified bolometric magnitude due to the presence of the secondary (Fig. \ref{fig:fwgmodified}). 
The modified $M_{\rm bol}$ is calculated by combining the V band flux of the primary and secondary and applying the BC based on the log $g$ and $T_{\rm eff}$ of the primary. The values of the BCs are taken from \cite{2011ApJS..193....1W}.

Comparing the tracks with modified and unmodified $M_{\rm bol}$, the increased flux from the secondary has the systematic effect of raising the tracks in the FGLR plane (Fig. \ref{fig:fwgmodified}). Because $F_{V, \, \mathrm{pri}}/F_{V, \, \mathrm{sec}}$ varies during the BSG stage (Fig. \ref{fig:fvratios}), the effect of the secondary on the modified $M_{\rm bol}$ also varies. The effect of the secondary on $M_{\rm bol}$ is more pronounced when the primary is a B-type supergiant than when it is an A-type supergiant.
When taking into account the binary fraction and the distribution of mass ratios, additional flux from secondary stars will produce a natural scatter in the FGLR plane. This may be some of the source of the observed scatter in the FGLR. Similar results are obtained for different orbital periods.
Although this natural scatter is still within the error bars obtained from observations and remains consistent with the observed scatter, it is a systematic increase in bolometric luminosity which could be important when using the FGLR to determine distances. Of course, if the calibration of the FGLR is equally affected, then the systematic effect would not affect distance determinations. It is important to note that Fig. \ref{fig:fwgmodified} represents the maximum effect and we expect a smaller effect for smaller mass ratios. 
%In order to properly quantify the systematic effect, population synthesis is required.

\subsection{Identifying BSGs in binary systems}
One way to deconstruct a composite spectrum from an unresolved binary system containing a BSG is through radial velocity measurements. The orbital velocities expected in the binary systems are indicated in Table \ref{table:1}. From Table \ref{table:1}, we see that the maximum amplitude of the orbital velocity is between 40 and 60 km s$^{-1}$, therefore only a fraction of these systems could be detected to be composite. The extragalactic FGLR studies work with a resolution of 5 {\AA} and will therefore not be able to identify most binaries through radial velocity variations.
In addition, some systems will present eclipses and thus might be distinguished through their photometric variability. We note, however, that for many of these systems the period is of the order of one year or more thus implying longer term observing campaigns.

\begin{figure}
	\centering
	\includegraphics[width=\hsize]{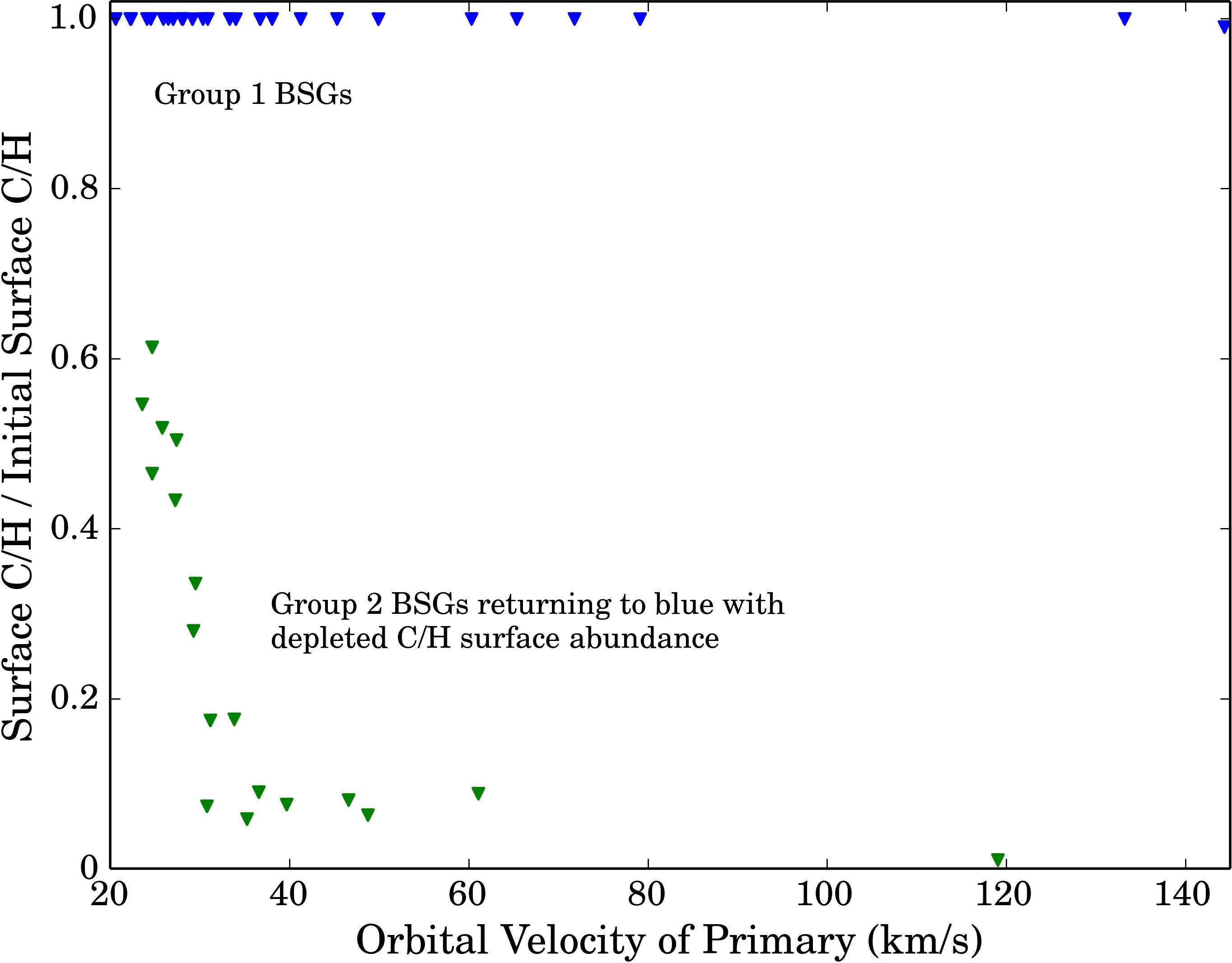}
	\caption{C/H surface abundance vs. orbital velocity for primary stars in our MESA binary models. The values were computed at a temperature of $T_{\rm{eff}}$ = 12\,500~K (see Table \ref{table:1}). The blue and green points represent stars moving towards the red and towards the blue in the HR diagram respectively. Not all of the stars moving towards the blue are considered BSGs by our criteria, due to low hydrogen surface abundances.}
	\label{fig:ch_vel}
\end{figure}

Figure \ref{fig:ch_vel} shows the expected trend between the C/H surface abundance and the orbital velocity for primary stars at $T_{\rm{eff}}$ = 12\,500~K (see Table \ref{table:1}).
The group of stars located at the top of the diagram have a similar C/H surface abundance to their initial value. These correspond to the stars evolving from the MS towards the RSG stage. They show no C depletion at the surface. A second group of stars show lower C/H surface abundances, indicating depletion of C at the surface. These stars are evolving towards the blue after experiencing strong mass loss in the RSG stage. The mass loss reveals the inner layers of the star which are depleted in C due to CNO processes. We see a trend of decreasing C/H surface abundances with increasing orbital velocity. Stars with smaller initial periods (and hence higher orbital velocities) undergo stronger mass loss (due to mass transfer) in the RSG stage, revealing deeper, more C depleted layers in the star. This trend, combined with radial velocities, could be used to identify blue stars that exist in binary systems. We note, however, that rotational mixing could affect this trend. Not all of these blue stars will necessarily be a BSG, as some may show strongly depleted surface hydrogen. To be tested, the two spectra should be distinguishable and relatively high resolution is needed to infer the radial velocity. Due to the high resolution required, this is more promising for nearby BSGs as observations of extragalactic BSGs typically use a spectral resolution of about 1000.

\subsection{Further Work} \label{sec:further}
Based on the results of this paper, we have determined that most BSGs in close binaries follow the observed FGLR and that the presence of unresolved secondary stars does not significantly affect the value of $M_{\rm bol}$, log $g$ or $T_{\rm eff}$ measured for primary BSGs.
We find that BSGs that form after mass transfer episodes (2nd or 3rd stage BSGs) are, in general, not consistent with the observed FGLR. 
As these stars have lost a substantial amount of mass, they may have in different wind properties. For instance, they may have extended atmospheres and show significant Balmer line emission, which would make the star observationally recognised as a blue hypergiant or a luminous blue variable (LBV). A similar effect has been found for single stars of 20 -- 25 $M_{\odot}$ by \citet{2013A&A...550L...7G}. We will explore the spectroscopic evolution of binary systems in a forthcoming paper.

We used the MLT++ (mixing length theory) scheme \citep{2013ApJS..208....4P} to avoid numerical issues. It remains to be seen whether this implementation is valid over the mass range we explored. Recently \citet{2018ApJ...853...79C} investigated the impact of the MLT++ on post-MS stars, in particular RSGs. Based on the results in that paper, it is possible that using MLT++ may affect the post mass transfer effective temperatures. The treatment of radiation-dominated envelopes in 1-D stellar evolution models is challenging and we expect that future models would benefit from improved physical implementation of radiation dominated envelopes.

Population synthesis calculations are needed to more precisely compare the predictions from these binary stellar evolution models with the observed FGLR. This would properly take into account the distribution of initial periods and mass ratios.
It is worth noting that \cite{2017PASA...34...58E} present a population synthesis based on the BPASS models to study primary BSGs in binary systems. Given the difference we have discussed between the BPASS and MESA models, it would be warranted to do similar work with the MESA models. While a full population synthesis of the same scale as BPASS would take significant time to complete, creating the necessary grid to focus on the FGLR is a feasible next step. We note that MESA has recently been used to do population synthesis of single stars \citep{2016ApJ...823..102C}. As we discuss above, rotation may significantly impact close binary systems and should be accounted for in future models.

\section{Conclusions} \label{sec:conclusions}

%In the first part of this work, we compared the position of BSGs in close binary systems in the $M_{\rm bol}$ versus Log ($g/T_{\rm eff}^4$) plane, assuming that the stars can be observed as individual objects. We then studied 
In this paper we investigated how binary evolution affects the properties of blue supergiants. In particular, we explored the effects on the bolometric luminosity and flux-weighted gravity (g/$T_{\rm eff}^4$). We initially assumed that the stars can be observed as individual objects, and then examined the impact of an unresolved secondary star on the spectrum of a primary BSG and the implications for the FGLR. Our main results and their implications are summarised below.

\begin{enumerate}

\item Based on the BPASS suite and a grid of models we computed with MESA, we find that most BSGs in close binary systems follow the observed FGLR. This is encouraging as it means that the FGLR is robust not only with respect to changes in the mass, metallicity and rotation but also with respect to multiplicity.

\item Our models indicate the possibility that there are some BSGs outside the FGLR observed scatter. These are produced when primary stars in a binary system undergo a mass transfer episode during the RSG stage and evolve back to the blue with a greatly reduced mass, and hence reduced flux-weighted gravity $g_{\rm F}$. Such systems may actually also be produced by single star evolution with strong mass losses during the RSG phase. In their spectroscopic FGLR studies, \citet{2008ApJ...681..269K} and \citep{2009ApJ...704.1120U} each found one such object in the galaxies NGC 300 and M33 respectively.
%A few such objects have been observed by \citet{2009ApJ...704.1120U}.

\item We estimate the frequency of these systems to be between 1 and 24\% depending on, among other factors, the surface fraction of hydrogen that an evolved blue star can have to still be considered a BSG. These percentages were estimated under the assumption that all stars exist in binary systems with periods between 1.4 and 3,000 days. 
If we take into account the existence of longer period binary systems and single stars, these percentages will decrease by an amount depending on the binary fraction and the initial period distribution of binary systems. The observations suggest a very small fraction of such objects. For nearby galaxies with distances smaller than 2 Mpc (M33, WLM, NGC3109, NGC300), only two objects out of a total of 81 (or 2.5\%) were found to be low mass FGLR outliers. However, we note that the target selection for spectroscopic FGLR distance determinations is heavily biased towards brighter objects to enable spectroscopy with a decent signal-to-noise. To assess the statistical effect of this selection bias on the fraction of outliers is difficult without population synthesis.

%\textbf{The observations suggest a very small fraction of such objects. For nearby galaxies with distances smaller than 2 Mpc (M33, WLM, NGC3109, NGC300), only two objects out of a total of 81 (or 2.5\%) where found to be low mass FGLR outliers. However, we note that the target selection for spectroscopic FGLR distance determinations is heavily biased towards brighter objects to enable spectroscopy with a decent signal-to-noise. To assess the statistical effect of this selection bias on the fraction of outliers is difficult without population synthesis.}
%While a small number of such FGLR outliers have been detected so far, a complete observational sample of blue stars would be needed to compare these models to observations.
%The small scatter of the FGLR can be used to infer that the scenarios leading to such outliers should be non-existent or very rare. In principle, a complete observational sample of blue stars would be needed to compare these models to observations.

\item In the context of extragalactic observations, the systems studied here would be unresolved by a 10 m telescope. Therefore, we studied  the impact of the presence of a secondary on the inferred magnitude, $B-V$ colour, log $g$ and effective temperature of a primary BSG in a binary system with a mass ratio of $q$ = 0.9. A high mass ratio will, in general, maximise the effects of the secondary on the primary spectrum. We find that, for a mass ratio of $q$ = 0.9, the contribution of a secondary star to the spectrum of a primary BSG has only a very small effect on the determination of $T_{\rm eff}$ and log g. The effects on the determination of interstellar reddening and bolometric magnitude are also small but systematic in the sense that the brightness will be overestimated by a few hundredths of a magnitude. In addition to this systematic effect, a natural scatter may be introduced to the FGLR. Detailed population synthesis calculations are needed to investigate this effect.

\item Our models suggest that some outliers to the FGLR could be the product of binary evolution. These come from systems with periods 2.8 < log($P$/days) < 3.5 as these systems produce BSGs returning from the RSG stage after a mass transfer episode. Interestingly, all these outliers would present strongly depleted C surface abundances.
\end{enumerate}

%In conclusion, we find that most BSGs in close binary systems follow the observed FGLR, although some possible outliers exist. The contribution of flux from an unresolved secondary has both a systematic effect on the FGLR and also produces a natural scatter. It is possible that this is one of the causes of the observed scatter in the FGLR however population synthesis is required to properly quantify this effect.

In conclusion, we find that most BSGs in close binary systems should be suitable for extragalactic distance determinations using the flux-weighted gravity luminosity relationship, although some possible outliers exist. The contribution of flux from an unresolved secondary has small systematic effect on the FGLR and also produces a natural scatter in the relationship.

After mass transfer and interaction with the companion, our results indicate that massive stars may only be recognised as blue supergiants in binary systems with a certain range of orbital periods, which depends on the mass ratio of the two components. For shorter orbital periods, different post-interaction spectra could be produced as the surface H abundances are significantly reduced compared to normal BSGs. These post-interacting systems could be observationally classified as blue hypergiants or LBVs. During or shortly after the mass transfer, the companion stars could be recognised as B[e] stars given the dense circumstellar medium that could be produced as a result of high mass-transfer rates, possibly coupled with the spin up of these companion stars by mass accretion. These evolutionary connections illustrate that the properties of blue supergiants evolving in binary systems analysed in this paper, and how they are connected to other classes of massive stars and supernova progenitors, remain an important topic for further exploration.

\begin{acknowledgements}
This work was partially carried out in Geneva Observatory. EF and JHG acknowledge support from the Irish Research Council New Foundations Award 206086.14414 `Physics of Supernovae and Stars'.
This work was also partially supported by the Swiss SNF project Interacting Stars, number 200020-172505. SCY was supported by the Korea Astronomy and Space Science Institute under the R \& D program supervised by the Ministry of Science and ICT.
We thank John Hillier for making CMFGEN available to the community and continuous support, and Bill Paxton and the MESA developers for making it publicly available. This work made use of v2.1 of the Binary Population and Spectral Synthesis (BPASS) models as last described in Eldridge, Stanway et al. (2017) and v10000 of MESA (Modules for Experiments in Stellar Astrophysics) as described in Paxton et al. (2015, 2013).
\end{acknowledgements}

\bibliographystyle{aa} 
\bibliography{refList}

\end{document}